\documentclass[10pt,twocolumn,letterpaper]{article}

\usepackage[pagenumbers]{cvpr} 

\definecolor{cvprblue}{rgb}{0.21,0.49,0.74}
\usepackage[pagebackref,breaklinks,colorlinks,allcolors=cvprblue]{hyperref}
\usepackage{booktabs,multirow,makecell,array}
\definecolor{lightyellow}{RGB}{255, 255, 204}
\usepackage[table]{xcolor}
\usepackage{algorithm}
\usepackage{algpseudocode}
\usepackage{amsmath}
\usepackage{diagbox}
\usepackage{tikz} 
\usepackage{pifont}
\newcommand{\redxmark}{{\color{red} \ding{55}}} 
\newcommand{\greencheck}{{\color{ForestGreen} \ding{51}}} %
\newcommand\blfootnote[1]{%
  \begingroup
  \renewcommand\thefootnote{}\footnote{#1}%
  \addtocounter{footnote}{-1}%
  \endgroup
}

\def\paperID{7374} 
\def\confName{CVPR}
\def\confYear{2026}

\title{Training-free, Perceptually Consistent Low-Resolution Previews with High-Resolution Image for Efficient Workflows of Diffusion Models}

\author{Wongi Jeong$^{*1}$ \qquad Hoigi Seo$^{*1}$  \qquad Se Young Chun$^{1,2\dag}$ \\
$^1$Dept. of Electrical and Computer Engineering, $^2$INMC \&  IPAI \\
Seoul National University, Republic of Korea \\
{\tt\small \{wg7139, seohoiki3215, sychun\}@snu.ac.kr}
}

\begin{document}
\maketitle
\begin{abstract}
Image generative models have become indispensable tools to yield exquisite high-resolution (HR) images for everyone, ranging from general users to professional designers. However, a desired outcome often requires generating a large number of HR images with different prompts and seeds, resulting in high computational cost for both users and service providers. Generating low-resolution (LR) images first could alleviate computational burden, but it is not straightforward how to generate LR images that are perceptually consistent with their HR counterparts. Here, we consider the task of generating high-fidelity LR images, called Previews, that preserve perceptual similarity of their HR counterparts for an efficient workflow, allowing users to identify promising candidates before generating the final HR image. We propose the commutator-zero condition to ensure the LR-HR perceptual consistency for flow matching models, leading to the proposed training-free solution with downsampling matrix selection and commutator-zero guidance. Extensive experiments show that our method can generate LR images with up to 33\% computation reduction while maintaining HR perceptual consistency. When combined with existing acceleration techniques, our method achieves up to 3$\times$ speedup. Moreover, our formulation can be extended to image manipulations, such as warping and translation, demonstrating its generalizability.
\end{abstract}
\blfootnote{* Authors contributed equally. $\dag$ Corresponding author.}
\section{Introduction}
Diffusion generative models~\cite{dhariwal2021diffusion, rombach2022high, esser2024scaling, ho2020denoising, song2021denoising}, with their strong ability to approximate complex probability distributions, have been used for diverse tasks, including image synthesis~\cite{rombach2022high, esser2024scaling, podell2024sdxl, chen2024pixart}, high-quality 3D asset generation~\cite{seo2023ditto, wang2023prolificdreamer, liu2023zero, shi2024mvdream}, video generation~\cite{wan2025wan, zheng2024open, ho2022video}, and text modeling~\cite{gong2023diffuseq, li2022diffusion, nie2025large}. Building on this foundation, flow matching models~\cite{liu2023flow, esser2024scaling, lipman2023flow, labs2025flux} that learn velocity fields have further expanded these capabilities, demonstrating improved scalability. Fueled by these advances, diffusion-based models have become powerful tools in real-world services, not only for creators such as designers and photographers but also general users in image and video synthesis.

However, since state-of-the-art generative AI tools produce diverse images from a single prompt, users require extensive seed (or prompt) searching to obtain the desired output. To speed up this workflow, several methods proposed to accelerate diffusion model inference. Caching-based approaches~\cite{chen2024delta, zou2024accelerating, liu2025reusing, ma2024deepcache, liu2025timestep}, reuse redundant features or predict future latents to enable faster synthesis, introduce memory overhead and may alter perceptual similarity between original images and generated images from acceleration. More recent methods~\cite{jeong2025upsample, tian2025training} achieve acceleration by spatially downsampling latents; however, this approach distorts the latent representation, hindering the recovery of high-quality images with composition and color fidelity.

\begin{figure*}[t]
  \centering
  \includegraphics[width=1.0\linewidth]{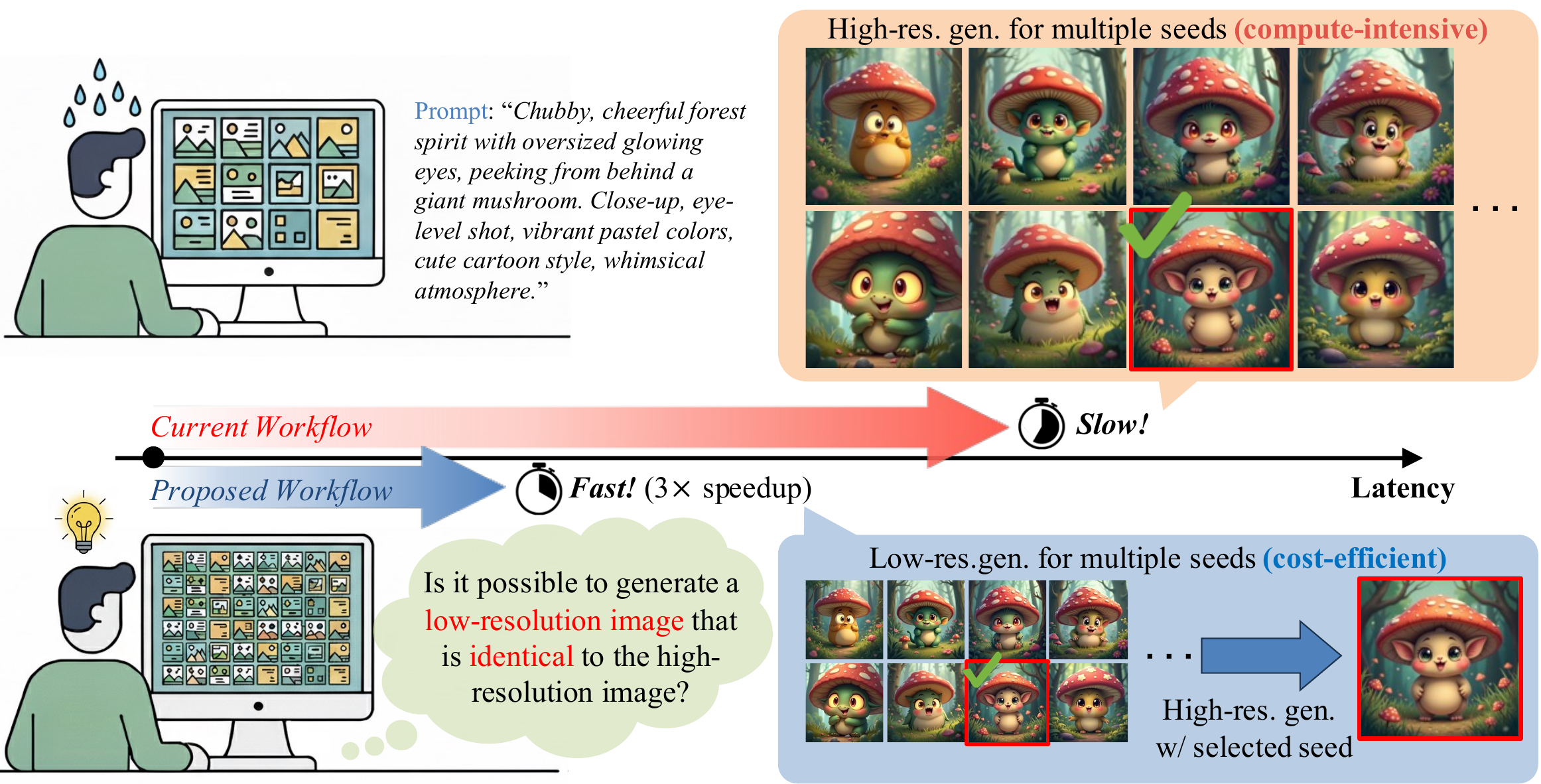}
  \vspace{-2em}
    \caption{\textbf{Motivation of Preview Generation.} Most users of generative AI usually produce diverse candidate images through repeated trials from multiple seeds (or prompts) to obtain a desired result. Our goal is to accelerate this workflow by generating low-resolution images that share the same content as their high-resolution counterparts but can be produced much faster in a \textbf{training-free} manner.}
    \label{fig:mask_selection}
\vspace{-1em}
\end{figure*}

We now raise a fundamental question: \textit{Instead of generating high-resolution images for every candidate, can we accelerate the seed search process by using perceptually similar low-resolution image candidates?}

In response to this question, we consider a workflow, called \textit{Preview Generation}, where users first generate diverse low-resolution (LR) candidates (Previews), then perform a high-resolution (HR) sampling using the selected seed and prompt. This enables efficient image selection and maintains quality with reduced computational cost. The core component for it is our proposed LR generation that preserves the color composition and structure of their HR counterparts. We reformulate this objective as enforcing the zero-vector commutator between the downsampling operation and the flow matching model, assuming trajectory compliance. To achieve this in a training-free manner, we introduce a principled selection of mutually exclusive downsampling operators that best satisfy the commutator-zero condition. A fixed-point–inspired commutator-zero guidance mitigates accumulated errors using only forward passes, eliminating costly backpropagation.

We extensively evaluate our method against various alternatives on state-of-the-art models, including FLUX.1-dev~\cite{frontier2024flux1dev} and Stable Diffusion 3.5-L~\cite{esser2024scaling} (SD3.5-L). Baselines include approaches that reduce the number of function evaluations (NFE), generate images directly in LR, or na\"ively downsample latents at the desired timestep. Across image quality, perceptual, and low-level similarity metrics, our method consistently yields LR images that more faithfully align with their HR counterparts. Furthermore, we demonstrate that the commutator-zero condition generalizes beyond downsampling, enabling integration into sampling processes involving operations such as warping and translation, underscoring the generalizability of our formulation.
Our contribution can be summarized as follows.
\begin{itemize}
    \item We consider \emph{Preview Generation}, the task to generate LR images with perceptually similar HR contents for efficient workflows and formulate this as the commutator-zero condition in the flow-matching model. We propose downsample matrix selection and commutator-zero guidance to enforce this condition in a training-free manner.
    \item Our extensive experiments across diverse metrics demonstrate that LR images from our approach outperform alternative solutions in perceptual similarity to the original HR images, achieving up to $1.53\times$ acceleration.
    \item We show that integrating the proposed method with temporal-axis acceleration yields a 3.05$\times$ speedup, and further demonstrate its strong generalization to a variety of image manipulation tasks.
\end{itemize}

\section{Related Works}
\paragraph{Flow matching models.}
Flow matching~\cite{lipman2023flow} is a generative framework that models the transformation of a simple prior distribution (\textit{e.g.}, Gaussian) into the complex data distribution (\textit{e.g.}, image) by learning a deterministic transport map via a continuous-time ODE. The evolution of a sample $\mathbf{x}_t$ over time is governed by the learned velocity field $v_\theta$:
\begin{equation}\label{eq:ode_flow}
dx_t=v_\theta(x_t, t)dt.
\end{equation}
This approach eliminates the need for stochastic sampling required by previous approaches. In particular, the conditional flow matching objective trains the neural network $v_\theta(x_t, t)$ to predict a target conditional velocity field. Rectified Flow~\cite{liu2023flow}, an efficient variant, uses a linear interpolation path between noise $x_0$ and the data sample $x_1$:
\begin{equation}\label{eq:rectified_flow_final}
x_t = (1 - t)x_0 + tx_1,\; x_0 \sim\mathcal{N}(0, \mathbf{I}), \; t \in [0, 1].
\end{equation}
This interpolated linear (rectified) path corresponds to a constant ground-truth conditional velocity field:
\begin{equation}
    u_t(x_t|x_1) = \frac{dx_t}{dt} = x_1 - x_0.
\end{equation}
The model is then trained to predict this velocity by minimizing the mean squared error between the predicted velocity $v_\theta(x_t, t)$ and $u_t(x_t|x_1)$, producing straight trajectories and thus requiring fewer function evaluations for sampling.

\paragraph{Temporal-axis acceleration.}
To reduce the high inference latency of large-scale flow matching models, several methods cache intermediate features~\cite{chen2024delta, zou2024accelerating, liu2025reusing, ma2024deepcache, liu2025timestep} for acceleration. Rectified flow-based models~\cite{esser2024scaling, liu2023flow, batifol2025flux, li2025omniflow, kong2024hunyuanvideo} with linear trajectories and Diffusion Transformers~\cite{peebles2023scalable} with inherent redundancy are primary targets. $\Delta$-DiT~\cite{chen2024delta} caches layer-wise feature residuals and skips redundant layers, ToCa~\cite{zou2024accelerating} applies token-wise caching, and TaylorSeer~\cite{liu2025reusing} leverages Taylor expansion to predict trajectories and update cached features. However, these methods yield only linear speedup with memory overhead proportional to cached features or timesteps, limiting scalability.

\paragraph{Spatial-axis acceleration.}
Orthogonal to caching-based acceleration, several methods reduce the number of tokens~\cite{tian2025training, jeong2025upsample, zhang2025training} for noise prediction, achieving quadratic speedup. Bottleneck Sampling~\cite{tian2025training} performs full-resolution denoising to capture global semantics, downsamples mid-stage latents to save computation, and upsamples later to restore quality. RALU~\cite{jeong2025upsample} begins with a low-resolution latent and selectively upsamples detailed regions such as edges for faster sampling. However, direct latent manipulation disturbs content preservation, causing deviations in composition and color; thus, accelerated outputs cannot faithfully reconstruct the HR counterpart.

\section{Method}
In diffusion models, the early denoising stages primarily define an image’s global layout~\cite{yang2023diffusion, kim2023leveraging, huang2024blue}. Building on this observation, we apply downsampling at a specific timestep $t_D$ during high-resolution (HR) sampling, leveraging the globally coherent representations formed early in the process to generate low-resolution (LR) images that remain closely aligned with their HR counterparts. This section motivates Preview generation (\cref{sec:sr_vs_hr}), introduces methods to address discrepancies between LR and HR images after downsampling (\cref{sec:compl_and_comm}, \cref{sec:method_D}, \cref{sec:commut}), and presents the overall algorithm (\cref{sec:algorithm}).
\begin{figure}[t]
  \centering
  \includegraphics[width=1.0\linewidth]{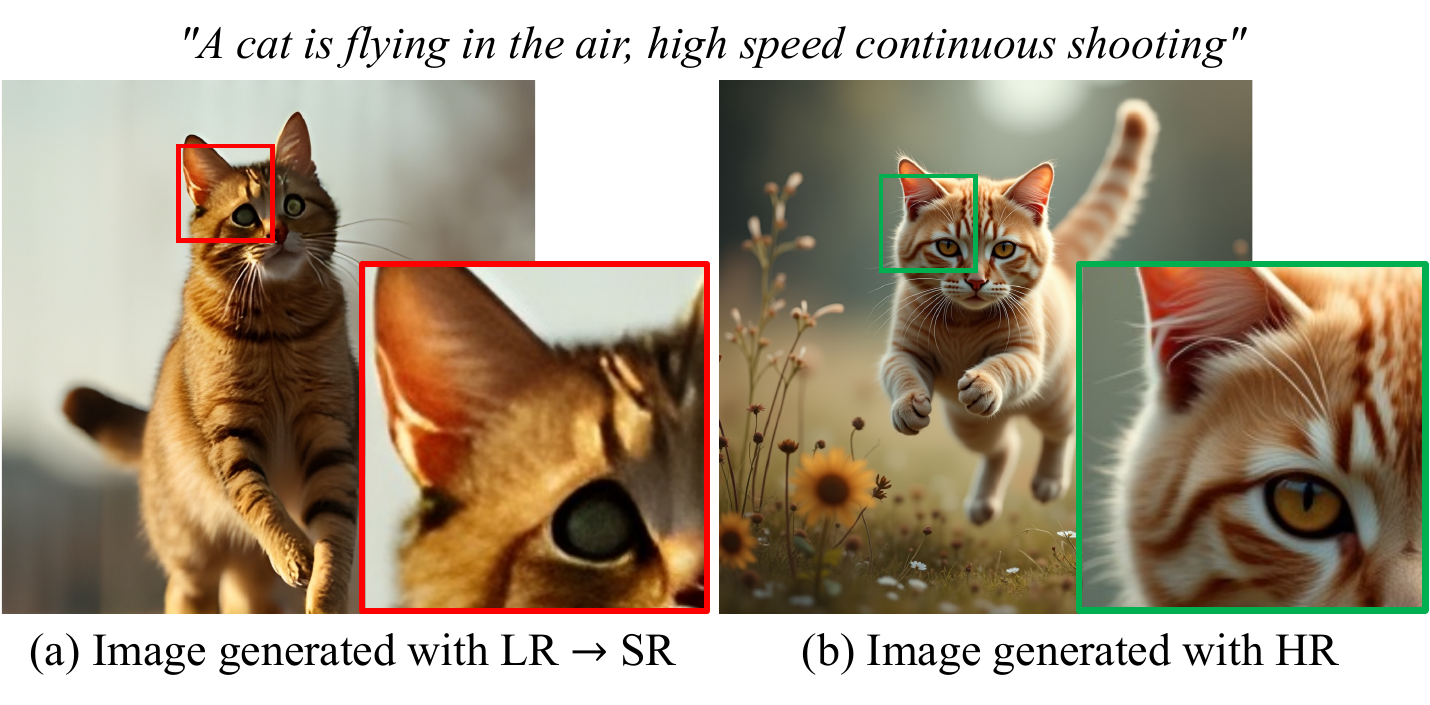}
  \vspace{-2em}
    \caption
    {\textbf{Comparison between SR-upsampled and directly generated HR images.} Using FLUX.1-dev, we \textbf{(a)} generated a low-resolution (LR, $256\times256$) image followed by $4\times$ super-resolution (SR) to obtain a high-resolution (HR, $1024\times1024$) image, and \textbf{(b)} directly generated an HR ($1024\times1024$) image. As shown in the close-up view, (a) fails to recover eye-region and fine fur details lost at the LR stage, whereas (b) preserves both global structure and fine details with high fidelity.}
    \label{fig:hr_vs_sr}
\vspace{-1em}
\end{figure}
\subsection{Why not super-resolution?}\label{sec:sr_vs_hr}
A natural question to our pipeline—first generating Previews for candidate selection and then synthesizing full-resolution images—is: \textit{why not generate LR images and upsample a chosen image via super-resolution?} Prior works on high-resolution synthesis~\cite{kim2024beyondscene, du2024demofusion, he2023scalecrafter, zhang2025diffusion, jeong2025latent} consistently show that errors introduced at LR propagate during upsampling.
As illustrated in Fig.~\ref{fig:hr_vs_sr}, the LR-upsampled image (left) lacks the fine details present in the directly generated HR image (right), consistent with previous findings.

\begin{figure*}[!t]
  \centering
  \includegraphics[width=1.0\linewidth]{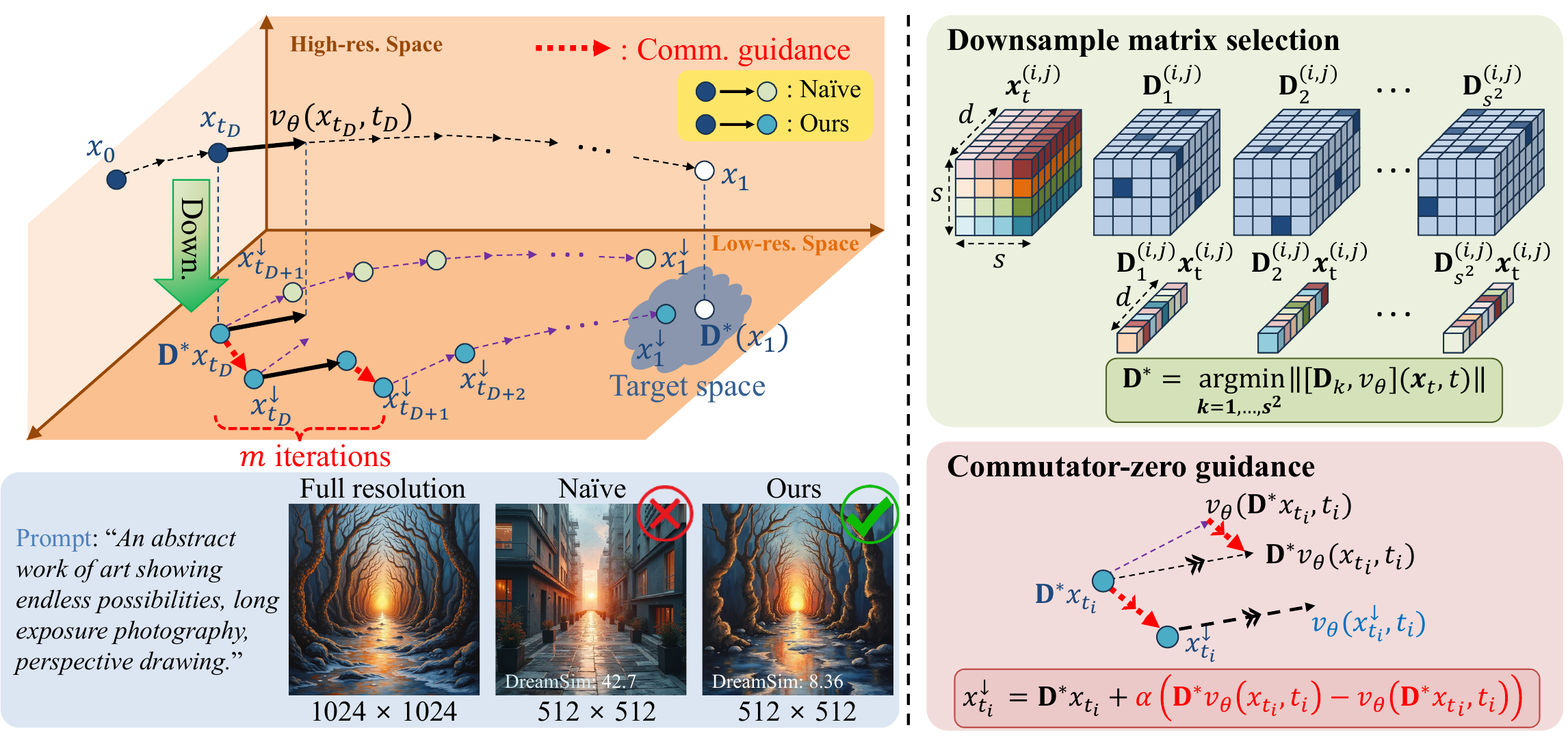}
  \vspace{-2em}
    \caption{\textbf{Overall framework.} \textbf{(Left, Top)} Overview of our proposed framework. Sampling is first performed in the high-resolution (HR) space up to timestep $t_D$, after which downsampling is applied using the selected downsample matrix from \textbf{(Right, Top)}. To maintain alignment with HR sampling, commutator-zero guidance is applied as shown in \textbf{(Right, Bottom)}. Finally, as illustrated in \textbf{(Left, Bottom)}, our method produces a low-resolution image with lower DreamSim score, which indicates better LR-HR consistency.}
    \label{fig:overall_framework}
    \vspace{-1em}
\end{figure*}

\subsection{Compliance and commutative-ness}\label{sec:compl_and_comm}
\paragraph{Compliance of the trajectory.}
To assess the feasibility of producing an LR image identical to the downsampled version of its HR counterpart, we define a flow ODE over the downsampled trajectory $x^\downarrow_t$ as follows:
\begin{equation}
dx^\downarrow_t = \mathbf{D}v_\theta(x_t, t)dt,
\end{equation}
where $\mathbf{D}\in\mathbb{R}^{\frac{hw}{s^2}\times hw}$ is a downsampling matrix with height $h$, width $w$, and scale factor $s$. Let $x^\downarrow_1$ and $x_1$ denote the final LR and HR images obtained from the trajectories ${x^\downarrow_t}$ and ${x_t}$, respectively. For the Preview generation task to be valid, the compliance condition $x^\downarrow_1 = \mathbf{D}x_1$ should hold. Although Zhang et al.~\cite{zhang2024flow} noted that this condition is not strictly satisfied in learned flow matching models, they reported strong empirical performance under this assumption. Similarly, we observe that assuming compliance for the downsampling operator yields consistent and effective results.

\begin{table}[!b]
    \caption{\textbf{Mean L2-norm of the commutator across models.} We report the spatial-wise averaged L2-norm of the commutator over 100 samples for FLUX.1-dev and Stable Diffusion 3.5-Large (SD3.5-L). All models generate images or videos with 30 function evaluations (NFE). These results illustrate that flow matching models do not satisfy the commutator condition.}
    \label{tab:commutator_norm}
    \centering
    \resizebox{\linewidth}{!}{
    \begin{tabular}{l|cc}
    \toprule
        Base model & FLUX.1-dev & SD3.5-L  \\
    \midrule
        $\|[\mathbf{D}, v_\theta](x_t, t)\|\; (\pm 1\sigma)$ & 111.03 $(\pm29.63)$ & 105.90 $(\pm22.38)$\\
    \bottomrule
    \end{tabular}
    }
    \vspace{-1em}
\end{table}

\paragraph{Commutative-ness of flow-matching.}
We aim to quickly synthesize the downsampled image that preserves the content of the original high-resolution image. To this end, we formulate to accelerate the sampling by feeding the learned flow matching model $v_\theta$ with the downsampled latent $\mathbf{D}x_t$ in place of the high-resolution latent $x_t$. However, this substitution raises a critical issue: Does the following commutator condition hold?
\begin{equation}\label{eq:commutator}
 [\mathbf{D}, v_\theta](x_t, t) \triangleq \mathbf{D} v_\theta(x_t, t)- v_\theta( \mathbf{D}x_t, t)\stackrel{?}{=} \mathbf{0}.
\end{equation}
In general, this \emph{commutator-zero condition} does not hold as seen in \cref{tab:commutator_norm}. Thus, we propose a method that encourages to minimize the norm of the commutator $\|[\mathbf{D}, v_\theta](x_t, t)\|$ with respect to 
the elements under our control, $\mathbf{D}$ and $x_t$. We design our approach from both $\mathbf{D}$ (\cref{sec:method_D}) and $x_t$ (\cref{sec:commut}) to better satisfy this condition in \cref{eq:commutator}.

\subsection{Downsample matrix selection}\label{sec:method_D}
We found that an appropriate choice of the downsampling matrix $\mathbf{D}$ alone reduces $\|[\mathbf{D}, v_\theta](x_t, t)\|$, improving alignment with HR images. However, this raises two issues: non-binary $\mathbf{D}$ may cause correlation of noise thereby degrading fidelity, and optimizing $\mathbf{D}$ is computationally expensive, hindering fast LR generation. To overcome these challenges, we propose generating multiple candidate $\mathbf{D}$ matrices and selecting the one that minimizes $\|[\mathbf{D}, v_\theta](x_t, t)\|$.

We generate candidates as follows. For each spatial block of size $s \times s$, we define a block-wise downsampling matrix $\mathbf{D}_{s\times s}:\mathbb{R}^{(s\times s)\times d}\rightarrow\mathbb{R}^{d}$, where $d$ is the channel dimension of the latent feature. One element is randomly selected from the $s^2$ spatial positions, yielding $s^2$ non-overlapping candidates. Aggregating the $hw/s^2$ block-wise matrices $\mathbf{D}_{s\times s}$ over the spatial domain forms the global downsampling operator $\mathbf{D}:\mathbb{R}^{(hw)\times d}\rightarrow\mathbb{R}^{(hw/s^2)\times d}$:
\begin{equation}
\mathbf{D}_k\triangleq
\Bigg(
   \bigoplus_{i=1}^{h/s}
   \;\bigoplus_{j=1}^{w/s}
   \mathbf{D}_{s\times s, k}^{(i,j)}
\Bigg)
\,\Pi, \; k\in \{1,\dots, s^2\},
\end{equation}
where $\Pi$ is a block-gathering permutation matrix that rearranges input latent rows so elements within each $s\times s$ block are contiguous. $\mathbf{D}_{s\times s}^{(i,j)}$ denotes the local downsampler for the $(i,j)$-th block, and $\oplus$ represents the direct sum forming a block-diagonal matrix of all local operators $\mathbf{D}_{s\times s}^{(i,j)}$. The $s^2$ mutually exclusive matrices constructed in this manner form a set $\mathcal{D}_\text{down}$ as below:
\begin{equation}
    \mathcal{D}_\text{down} \triangleq \{\mathbf{D}_1,\dots,\mathbf{D}_{s^2}\},\; \mathbf{D}_i\odot\mathbf{D}_j=\mathbf{0}\;(i\neq j).
\end{equation}
Next, for each element in $\mathcal{D}_\text{down}$, we calculate the commutator and select the matrix $\mathbf{D}_i$ with the lowest mean L2-norm of the commutator to get the final downsample matrix $\mathbf{D}^\ast$:
\begin{equation}\label{eq:select_downsample}
    \mathbf{D}^\ast = \arg\min_{i=1,\dots,s^2} \|[\mathbf{D}_i, v_{\theta}](x_t, t)\|, 
\quad\mathbf{D}_i\in \mathcal{D}_\text{down}.
\end{equation}
Through this approach, we can obtain $\mathbf{D}^\ast$ with the same computational cost as evaluating a single $v_\theta(x_t, t)$. Using $\mathbf{D}^\ast$, we define the downsampled latent of $x_t$ as follows:
\begin{equation}
    x^{\downarrow}_t \triangleq \mathbf{D}^\ast x_t.
\end{equation}

\subsection{Commutator-zero guidance}\label{sec:commut}
We now turn to $x^\downarrow_t$ manipulation to reduce the commutator norm. The most straightforward approach is to update $x_t$ via backpropagation; however, this strategy contradicts our objective of rapid LR image synthesis. Instead of relying on computation-heavy gradient updates and inspired by methods based on fixed-point iteration~\cite{hong2024gradient, bai2024fixed, hang2024exploring}, we adopt a fixed-point iteration–like update rule.
\begin{equation}
x^{\downarrow,k+1}_t=x^{\downarrow,k}_t+\alpha\cdot(\mathbf{D}^\ast v_\theta(x_t, t)-v_\theta(x^{\downarrow,k}_t, t)),
\end{equation}
where $x^{\downarrow,k}_t$ is the latent with the $k$-th iteration, and $\alpha$ is a step size. However, this update rule requires storing $x_t$ at every step $t$, and computing $v_\theta(x_t, t)$ is prohibitively slow. Even without storing $x_t$, approximating it through $x^{\downarrow,k}_t$ is inaccurate, and finding a suitable inverse of $\mathbf{D}^\ast$ for approximation is challenging. We thus turn our attention to the following well-known property of rectified flow~\cite{ke2025proreflow, yan2024perflow, bertrand2025closed}:
\begin{equation}\label{eq:v_approx}
    v_\theta(x_{t_0}, t) \approx v_\theta(x_{t_0+\Delta t}, t+\Delta t),
\end{equation}
where $\Delta t$ denotes a small deviation in timestep. \cref{eq:v_approx} indicates that, since the velocity field learned by the rectified flow is linearized, its values remain similar within the neighborhood of any timestep $t_0$. Motivated by this property, we propose to reuse the previously computed $v_\theta(x_{t_D} t_D)$ at the downsampling timestep $t_D$ \textit{for free} instead of explicitly computing $v_\theta(x_t, t)$ at timestep $t$ as follows:
\begin{equation}\label{eq:final_correction}
x^{\downarrow,k+1}_t =x^{\downarrow,k}_t+\alpha\cdot(\underbrace{\mathbf{D}^\ast v_\theta(x_{t_D}, t_D)}_{\text{\cref{eq:v_approx}}}-v_\theta(x^{\downarrow,k}_t, t)).
\end{equation}
Through this approach, we achieve computational efficiency by updating $x^{\downarrow,k}_t$ using only $v_\theta(x^{\downarrow,k}_t, t)$, which is substantially less expensive to compute than $v_\theta(x^{k}_t, t)$.

Our implementation satisfies the condition in \cref{eq:v_approx} by restricting the updates to $m$ steps after $t_D$. For efficiency, we perform an update with $k=1$ at each of these timesteps.

\begin{algorithm}[t]
\caption{Preview generation with our method}
\label{alg:downsample}
\begin{algorithmic}[1]
\Require Rectified flow model $v_\theta$, NFE $N$, timestep schedule $t_i$, height $h$, width $w$, scale factor $s$, downsample step $D$ where $D/N \approx 0.3$, correction iteration $m$, correction step size $\alpha$.
\Ensure Denoised output $x_1$
\State $x_0 \sim \mathcal{N}(0, I)$ \Comment{$t_0=0$, $t_N=1$}
\For{$i = 0, \dots, N-1$}
    \If{$t_i=t_D$}
        \State $v_\theta(x_{t_D}, t_D) \gets v_\theta(x_{t_i}, t_i)$
        \For{$k = 0, \dots, s^2-1$}
            \State $\mathbf{D}_k\gets \Big(
               \bigoplus_{l=1}^{h/s}
               \;\bigoplus_{m=1}^{w/s}
               \mathbf{D}_{s\times s, j}^{(l,m)}
            \Big)$
        \EndFor
        \State $\mathbf{D}^\ast\gets \arg\min_{k=1,\dots,s^2} \|[\mathbf{D}_k, v_{\theta}](x_{t_i}, t_{t_i})\|$
        \Statex \hfill{\footnotesize\(\triangleright\) Downsample matrix selection (\cref{sec:method_D})}
        \State $x_{t_i} \gets \mathbf{D}^\ast x_{t_i}$
    \EndIf
    \If{$t_i \geq t_D$ \textbf{and} $t_i < t_{D+m+1}$ }
    \State $x_{t_i} \gets x_{t_i}+\alpha\cdot\big(\mathbf{D}^\ast v_\theta(x_{t_D}, t_D)-v_\theta(x_{t_i}, {t_i})\big)$
    \Statex \hfill{\footnotesize\(\triangleright\) commutator-zero guidance (\cref{sec:commut})}
    \EndIf
    \State $x_{t_{i+1}} \gets x_{t_i} +(t_i - t_{i+1})\cdot v_\theta(x_{t_i}, t_i)$
\EndFor
\State \Return $x_1$
\end{algorithmic}
\end{algorithm}

\begin{table*}[t]
    \caption{\textbf{Quantitative comparison on FLUX.1-dev and SD3.5-L.} Using the FLUX.1-dev and Stable Diffusion 3.5-Large (SD3.5-L) models with $\text{NFE}=30$, we generate HR ($1024\times1024$) reference images. We compare three variants: (i) reduced-NFE generation, (ii) LR ($512\times512$) generation with the same NFE, and (iii) a na\"ive baseline applying nearest downsampling at timestep $t_D$. Each method is evaluated for computational efficiency, image quality, and consistency with the reference. Numbers in parentheses denote the NFE used; if omitted, $\text{NFE}=30$ is applied as in the reference. We denote the best performance in \textbf{bold} and second-best performance with \underline{underline}.}
    \label{tab:quanti}
    \vspace{-0.5em}
    \centering
    \resizebox{\linewidth}{!}{
    \begin{tabular}{m{2.5cm}|ccc|c|cc|cc}
    \toprule
    \multirow{2}{*}{Method} & \multicolumn{3}{c|}{Efficiency} & Image quality & \multicolumn{2}{c|}{Perceptual sim.} & \multicolumn{2}{c}{Low-level sim.} \\
    & Latency(s)$\downarrow$ & TFLOPs$\downarrow$ & Speed$\uparrow$ & PIQE$\downarrow$ & DreamSim$\downarrow$ & DiffSim$\uparrow$ & PSNR (dB) $\uparrow$ & FSIM$\uparrow$ \\
    \midrule
    FLUX.1-dev \hspace*{\fill}\mbox{(30)} & 14.78 & 1800.49 & 1.00$\times$ & 28.92 & - & - & - & -  \\
    \midrule
    FLUX.1-dev \hspace*{\fill}\mbox{(20)} & 10.01 & 1205.25 & 1.49$\times$  & 34.78 & \:\:\underline{7.25} & \underline{0.8686} & \underline{20.633} & \textbf{0.8268} \\
    Low-res. & \:\:4.94 & \:\:602.96 & 2.99$\times$ & 
    \underline{29.11} & 21.02 & 0.7352 & 11.936 & 0.6147 \\
    Na\"ive Down. & \:\:8.53 & 1029.43 & 1.75$\times$ & 31.71 & \:\:9.20 & 0.8584 & 18.221 & 0.7375 \\
    \textbf{Ours} &\cellcolor{lightyellow}\:\:9.76 &\cellcolor{lightyellow}1178.30 &\cellcolor{lightyellow}1.53$\times$ &\cellcolor{lightyellow}\textbf{28.55} &\cellcolor{lightyellow}\:\:\textbf{6.83} &\cellcolor{lightyellow}\textbf{0.8721} &\cellcolor{lightyellow}\textbf{21.182} &\cellcolor{lightyellow}\underline{0.7953} \\
    \bottomrule
    \end{tabular}
    }
    \resizebox{\linewidth}{!}{
    \begin{tabular}{m{2.5cm}|ccc|c|cc|cc}
    \toprule
    \multirow{2}{*}{Method} & \multicolumn{3}{c|}{Efficiency} & Image quality & \multicolumn{2}{c|}{Perceptual sim.} & \multicolumn{2}{c}{Low-level sim.} \\
    & Latency(s)$\downarrow$ & TFLOPs$\downarrow$ & Speed$\uparrow$ & PIQE$\downarrow$ & DreamSim$\downarrow$ & DiffSim$\uparrow$ & PSNR (dB) $\uparrow$ & FSIM$\uparrow$ \\
    \midrule
    SD3.5-L \hspace*{\fill}\mbox{(30)}& 14.51 & 1533.81 & 1.00$\times$ & 29.39 & - & - & - & -  \\
    \midrule
    SD3.5-L \hspace*{\fill}\mbox{(20)} & \:\:9.87 & 1029.15 & 1.49$\times$ & 36.28 & 15.61 & 0.7893 & 13.496 & 0.6782 \\
    Low-res. & \:\:4.74 & \:\:531.83 & 2.88$\times$ & 31.95 & 28.46 & 0.6538 & \:\:8.318 & 0.5517 \\
    Na\"ive Down. & \:\:8.29 & \:\:890.58 & 1.72$\times$ & \textbf{29.75} & \underline{14.81} & \underline{0.7956} & \underline{13.858} & \underline{0.6919} \\
    \textbf{Ours} &\cellcolor{lightyellow}\:\:9.36 &\cellcolor{lightyellow}1019.17 &\cellcolor{lightyellow}1.50$\times$ &\cellcolor{lightyellow}\underline{31.55} &\cellcolor{lightyellow}\textbf{13.47} &\cellcolor{lightyellow}\textbf{0.8117}
    &\cellcolor{lightyellow}\textbf{14.457} &\cellcolor{lightyellow}\textbf{0.7408} \\
    \bottomrule
    \end{tabular}
    }
    \vspace{-1em}
\end{table*}

\subsection{Algorithm}\label{sec:algorithm}
We present the complete algorithm for the proposed method in \cref{alg:downsample}. As noted earlier, early steps primarily capture global structure. To ensure the LR image aligns with its HR counterpart, sampling is first performed at full resolution up to timestep $t_D$, after which the feature map is downsampled by a scale factor $s$. The full-resolution velocity predictions are stored for commutator-zero guidance. A set of downsampling matrix candidates is then generated, and the one $\mathbf{D}^*$ with the smallest commutator is selected. With $\mathbf{D}^*$ and the stored predictions, commutator-zero guidance is applied for the next $m$ timesteps, producing a LR image $x_1$ consistent with its HR counterpart.

\section{Experiments}
We evaluate the efficacy of our approach both quantitatively and qualitatively. All baseline methods are assessed with state-of-the-art flow matching models, FLUX.1-dev~\cite{frontier2024flux1dev} and Stable Diffusion 3.5-Large~\cite{esser2024scaling} (SD3.5-L) with BFloat16 precision. For hyperparameter settings, both models use $m=5$, NFE $N=30$, downsample step $D=10$, with $\alpha=0.04$ for FLUX and $\alpha=0.01$ for SD3.5.
\subsection{Quantitative results}
\paragraph{Baselines.}
We compare our method against three baselines: (i) reducing the number of function evaluations to $\text{NFE}=20$, (ii) directly generating lower-resolution ($512\times512$) images, and (iii) performing downsampling at timestep $t_D$ using a na\"ive strategy followed by continued generation.
\paragraph{Dataset and Metrics.}
To better reflect practical generation workflows, we evaluate our method on the PixArt-Eval30K~\cite{chen2024pixart} dataset, which features semantically detailed and artistically rich prompts describing complex and dynamic scenes, instead of the conventional MS-COCO dataset~\cite{lin2014microsoft}. For quantitative evaluation, 5K prompts are randomly selected from PixArt-Eval30K for evaluation.

To evaluate our approach, we employ standard similarity metrics, including PSNR and FSIM~\cite{zhang2011fsim}. Since seed selection mostly relies on perceptual similarity, we additionally use perceptual metrics DreamSim~\cite{fu2023dreamsim} and DiffSim~\cite{song2025diffsim}, commonly adopted for measuring perceptual correspondence~\cite{seo2025skrr, tewel2024training, jones2024customizing}. We also use PIQE~\cite{venkatanath2015blind}, a no-reference image quality assessment metric. All images were evaluated with the resolution of $512\times512$. Further details are provided in the supplementary materials.
\paragraph{Comparison with baselines.}
We present quantitative comparison results in \cref{tab:quanti}. On FLUX.1-dev, our method achieves the best performance in image quality, perceptual similarity, and PSNR, while ranking second in FSIM. Notably, the NFE-reduced alternative that scored highest in FSIM exhibits the lowest overall image quality. On SD3.5-L, our method attains the best performance across all metrics except for image quality, where it ranks second. Nevertheless, it surpasses the na\"ive baseline in both perceptual and low-level similarity, demonstrating the efficacy and generalizability of the proposed method.

\begin{table}
    \caption{\textbf{Quantitative comparison on FLUX.1-dev with temporal-axis acceleration.} We compare our approach combined with orthogonal temporal-axis acceleration. Our method achieves superior performance across all metrics, including image quality, perceptual similarity, and low-level similarity.}
    \label{tab:temporal_acceleration}
    \vspace{-0.5em}
    \centering
    \resizebox{\linewidth}{!}{
    \begin{tabular}{l|c|ccc}
    \toprule
    Method & Speed$\uparrow$ & PIQE$\downarrow$ & DreamSim$\downarrow$ & PSNR (dB)$\uparrow$ \\
    \midrule
    FLUX.1-dev (30) & $1.00\times$ & 28.92 & - & - \\
    \midrule
    FLUX.1-dev (10) & $2.95\times$ & 32.82 & 16.97 & 16.427\\
    Taylor only &$3.21\times$ & 28.10 & \:\:9.17 & 18.667 \\
    \textbf{Ours + Taylor} &\cellcolor{lightyellow}$3.05\times$ &\cellcolor{lightyellow}\textbf{27.47} &\cellcolor{lightyellow}\:\:\textbf{7.79} &\cellcolor{lightyellow}\textbf{19.953} \\
    \bottomrule
    \end{tabular}
    }
    \vspace{-1em}
\end{table}

\paragraph{Integration with temporal-axis acceleration.}
Since temporal-axis acceleration methods are orthogonal to our approach, they can be combined with our framework to achieve additional efficiency. As shown in \cref{tab:temporal_acceleration}, we integrate the state-of-the-art diffusion acceleration method, TaylorSeer~\cite{liu2025reusing}, with our approach. The combined method achieves $3.05\times$ speedup, higher HR consistency and image quality compared to using the same level of NFE reduction.

\begin{figure*}
    \centering
    \includegraphics[width=\linewidth]{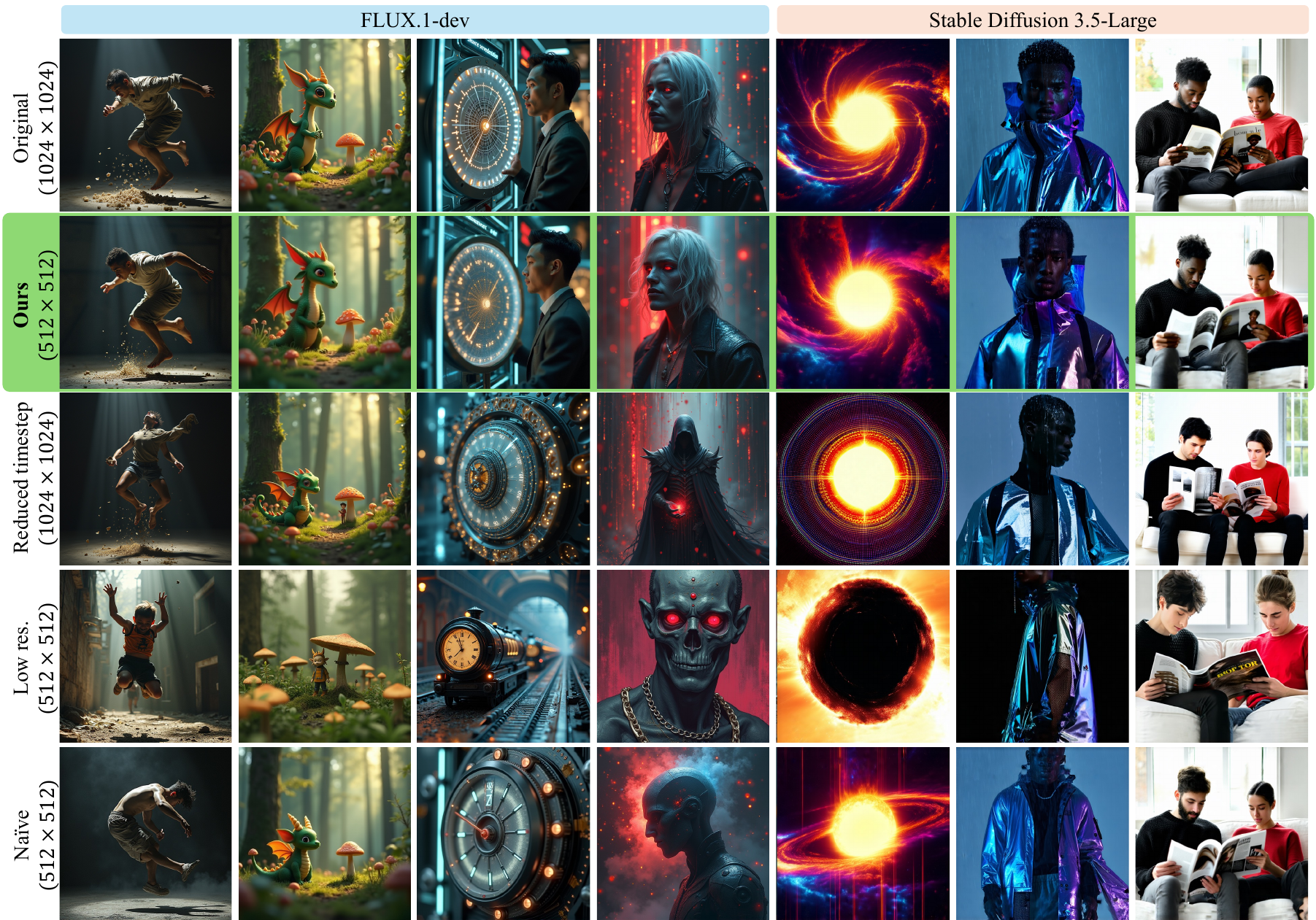}
    \vspace{-2em}
    \caption{\textbf{Qualitative comparison of our proposed method.} While other simple alternatives often result in changes to composition, object size, or even color tone, our proposed approach synthesizes low-resolution images faster while preserving the composition and color fidelity of the original image. The prompts used for image generation are provided in the supplementary materials.}
    \label{fig:main_quali}
    \vspace{-1em}
\end{figure*}

\begin{figure}
    \centering
    \includegraphics[width=\linewidth]{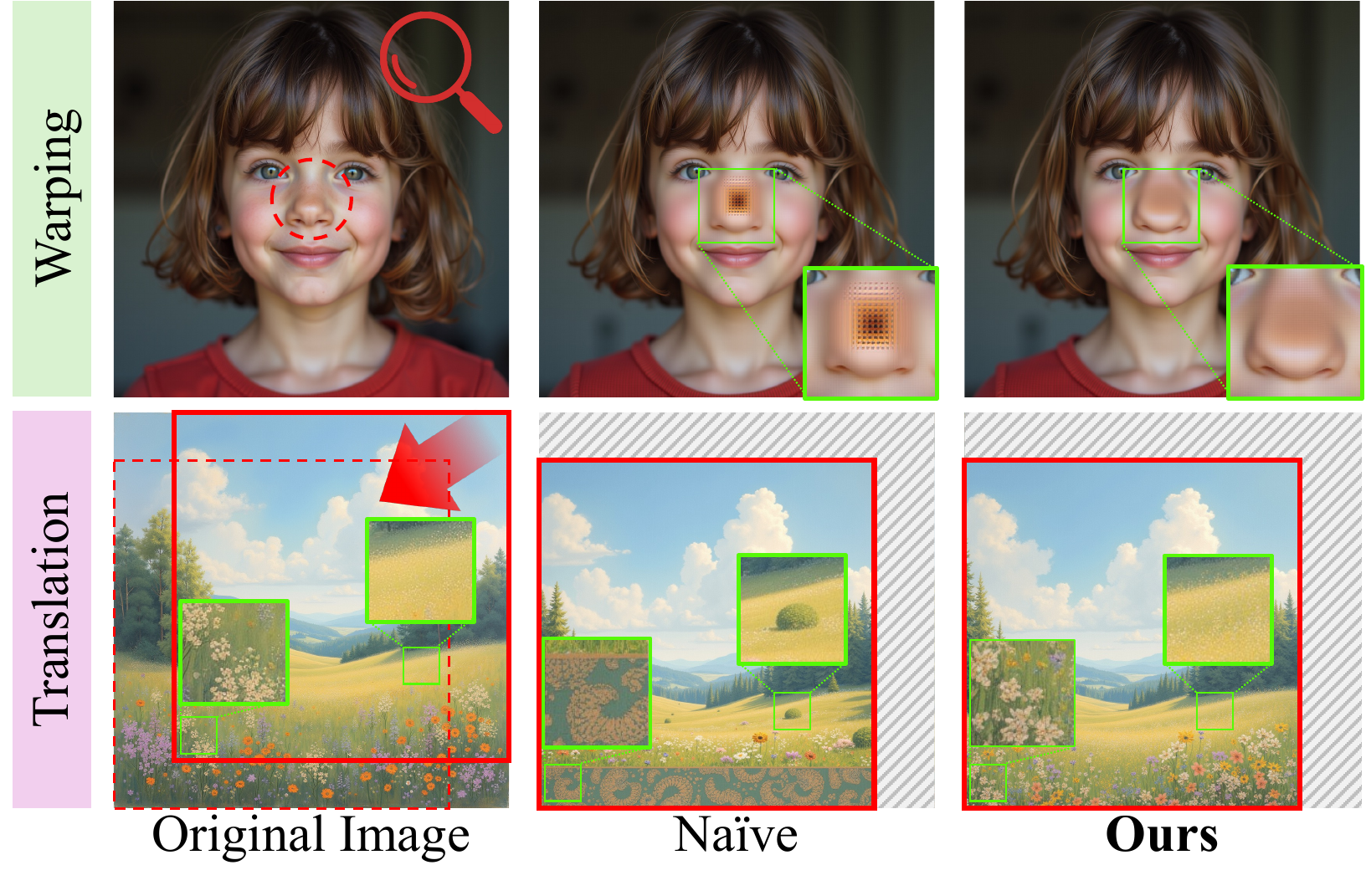}
    \caption{\textbf{Generalization of commutator-zero guidance.} We show that commutator-zero guidance can be expanded to other operations. For warping, a large kernel ($128\times128$) with correlation correction produces distortion, while our method effectively handles artifacts. For translation, na\"ive cause noticeable difference and unintended objects, whereas ours preserves image content.}
    \label{fig:generalization}
    \vspace{-1em}
\end{figure}

\subsection{Qualitative results}
In \cref{fig:main_quali}, we present qualitative comparisons for the $512\times512$ setting. While other alternatives often result in changes to image composition, lighting, or color scheme, the proposed strategy successfully produces low-resolution images that are perceptually very similar to their high-resolution counterparts. Additional qualitative results for this setting, generation at $256\times256$ resolution, and results combined with temporal-axis acceleration are provided in the supplementary materials.

\subsection{Generalization of commutator-zero guidance}
Our commutator-zero guidance implies that any operation that does not incur correlation between noise can substitute for $\mathbf{D}$. To verify this, we tested two image manipulations during synthesis: translation and warping. For translation, our method better preserves the original image, without generating additional objects or pattern that does not exists in an original image. Since warping incurs correlation, we adapted correction~\cite{tian2025training,jeong2025upsample}. As shown in \cref{fig:generalization}, na\"ive warping and translation introduces artifacts and unintended objects, whereas our approach effectively removes them.

\subsection{Ablation study}
We conducted an ablation study to assess the individual contributions of the downsampling matrix selection and commutator-zero guidance components, and evaluate the effectiveness of their alternative configurations. All experiments were performed on 500 randomly selected prompts from the PixArt-Eval30K dataset using the FLUX.1-dev model. Further studies are provided in the supplementary.
\paragraph{Effect of $\mathbf{D}$ selection.}
We validate our method by comparing strategies for selecting the downsampling matrix: spatial-wise Nearest, random sampling, and a variant using $\arg\max(\cdot)$ in \cref{eq:select_downsample}. As shown in \cref{tab:ablation_selection}, Nearest $\mathbf{D}$, which loses more information, performs worst. The $\arg\max$ variant also underperforms, while Random $\mathbf{D}$ lies between it and our $\arg\min$ approach. These results show that minimizing the commutator yields better performance.
\paragraph{Effect of commutator-zero guidance.}
To examine the effect of commutator-zero guidance, we conducted an ablation study as shown in \cref{tab:ablation_cc}. Without applying this correction, the HR counterpart alignment noticeably weakens, leading to a decrease in low-level similarity and perceptual consistency. These results underscore the validity of our formulation and the contribution of the commutator-zero guidance to the effectiveness of our method.
\paragraph{Hyperparameter study.}
We performed a hyperparameter study by varying the correction iteration count $m$ and step size $\alpha$ in the commutator-zero guidance. As shown in \cref{tab:ablation_hyper}, increasing $m$ consistently improves PSNR and produces outputs more aligned with the HR counterpart, validating the module’s effectiveness. We further observe that each $m$ has an optimal $\alpha$. Balancing performance and computational efficiency, we set $m=5$ and $\alpha=0.04$.

\begin{table}
    \caption{\textbf{Ablation study on $\mathbf{D}$ selection.} We compare three strategies for selecting $\mathbf{D}$: (i) nearest-neighbor downsampling, (ii) random sampling, and (iii) $\mathbf{D}^*$ obtained via the $\arg\max(\cdot)$ operation in Eq.~\ref{eq:select_downsample}. The results show that our approach, which minimizes the commutator value, achieves the best performance.}
    \label{tab:ablation_selection}
    \centering
    \resizebox{0.95\linewidth}{!}{
    \begin{tabular}{c|ccc}
    \toprule
         & PIQE$\downarrow$ & DreamSim$\downarrow$ & PSNR$\uparrow$ (dB) \\
    \midrule
    Nearest $\mathbf{D}$ & 29.72 & 10.39 & 18.069 \\
    Random $\mathbf{D}$ & 27.35 & 7.40 & 20.851\\
    $\arg\max(\cdot)$ (Eq.~\ref{eq:select_downsample}) & 27.83 & 8.55 & 20.408 \\
    \midrule
    Full \textbf{(Ours)} & \textbf{27.08} & \textbf{7.05} & \textbf{20.962}\\
    \bottomrule
    \end{tabular}
    }
\end{table}

\begin{table}
    \caption{\textbf{Ablation study on commutator-zero guidance.} Comparing results with and without commutator-zero guidance (CG) shows that while computational efficiency slightly decreases with correction, both perceptual similarity and PSNR increase, highlighting the effectiveness of our proposed method.}
    \label{tab:ablation_cc}
    \centering
    \resizebox{0.75\linewidth}{!}{
    \begin{tabular}{c|ccc}
    \toprule
      CG & PIQE$\downarrow$ & DreamSim$\downarrow$ & PSNR$\uparrow$ (dB) \\
    \midrule
    \redxmark & 31.89 & 8.56 & 19.115\\
    \greencheck & \textbf{27.08} & \textbf{7.05} & \textbf{20.962}\\
    \bottomrule
    \end{tabular}
    }
\end{table}

\begin{table}
    \caption{\textbf{Ablation study on hyperparameter $m$ and $\alpha$.} We investigate how varying $m$ and $\alpha$ affects performance. Increasing $m$ improves PSNR but also raises computational demand, showing a clear trade-off. Each $m$ exhibits an optimal step size $\alpha$.}
    \label{tab:ablation_hyper}
    \centering
    \resizebox{\linewidth}{!}{
    \begin{tabular}{c|cccccc}
    \toprule
        \diagbox{$m$}{$\alpha$} & 0.01 & 0.02 & 0.03 & \textbf{0.04} & 0.05 & 0.06\\
    \midrule
        3 & 19.788 & 20.256 & 20.522 & 20.645 & 20.599 & 20.393 \\
        4 & 19.961 & 20.546 & 20.859 & 20.930 & 20.804 & 20.551 \\
        \textbf{5} & 20.119 & 20.794 & 21.104 & \textbf{21.146} & 20.957 & 20.543\\
        6 & 20.260 & 21.004 & 21.303 & 21.293 & 21.004 & 20.569 \\
        7 & 20.387 & 21.182 & 21.467 & 21.389 & 21.025 & 20.503\\
    \bottomrule
    \end{tabular}
    }
\end{table}

\section{Discussion}
\begin{figure}[t]
    \centering
    \includegraphics[width=\linewidth]{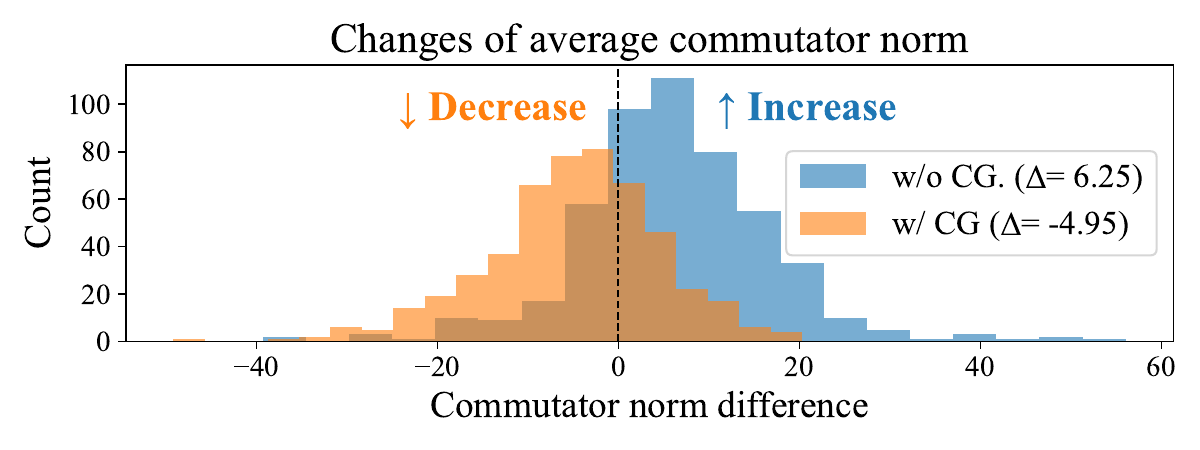}
    \vspace{-2em}
    \caption{\textbf{Changes in average commutator norm.} We measure the change in commutator norm at timesteps $t_D$ and $t_{D+m}$ with and without applying commutator-zero guidance (CG). When CG is applied, the commutator norm decreases significantly, whereas without CG, it shows a statistically significant increase.}
    \label{fig:cc_norm_change}
\end{figure}

\paragraph{Analysis on the norm of commutator.}
While downsample matrix selection explicitly minimizes the commutator, the effect of commutative-ness guidance (CG) requires empirical validation. As shown in \cref{fig:cc_norm_change}, CG effectively reduces the commutator magnitude, aligning the LR trajectory more closely with the HR counterpart. A Wilcoxon signed-rank test comparing commutator norms between $t_D$ and $t_{D+m}$ shows that without CG, the commutator increases significantly ($p=9.62\times10^{-38}$), whereas with CG it decreases significantly ($p=3.38\times10^{-25}$), confirming improvement at the 0.05 significance level.

\begin{figure}
    \centering
    \includegraphics[width=\linewidth]{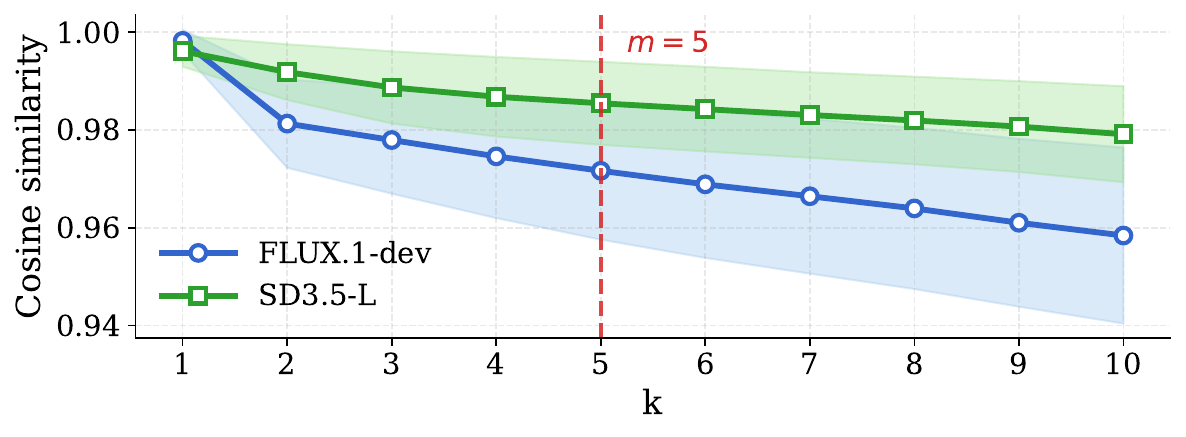}
    \vspace{-2em}
\caption{\textbf{Cosine similarity analysis on velocity consistency of flow-matching models.} We report the mean and standard deviation of cosine similarity between $v_\theta(x_{t_D}, t_D)$ and $v_\theta(x_{t_{D+k}}, t_{D+k})$ over 500 samples. The similarity remains above 0.95 for $k\le5$, supporting the validity of the approximation.}
    \label{fig:dis_cossim}
\end{figure}

\paragraph{Consistency of velocity field.}
As discussed in \cref{sec:commut}, we promote commutative-ness without extra computation using the approximation in \cref{eq:v_approx}. To validate its accuracy within our chosen $m$ iterations, we measured the cosine similarity between $v_\theta(x_{t_D}, t_D)$ and $v_\theta(x_{t_{D+k}}, t_{D+k})$ over 500 samples. The results, shown in \cref{fig:dis_cossim}, indicate that with $D=9$ and $m=5$, the similarity remains above 0.95, confirming the validity of the assumption in \cref{eq:v_approx}.

\section{Conclusion}
In this work, we revisited the long-standing efficiency challenge in diffusion-based generative models and introduced \textit{Preview Generation}, a framework that enables accelerated image synthesis workflow. By reformulating the problem as enforcing a commutator-zero condition between the downsampling operator and the flow matching model, we derived a principled means of ensuring that low-resolution (LR) trajectories remain consistent with their high-resolution (HR) counterparts. Our approach further integrates downsampling matrix selection and commutator-zero guidance, enabling faithful LR generation. Extensive experiments with various models validate the effectiveness of our method across diverse metrics. Beyond acceleration, our formulation generalizes to spatial manipulations, suggesting a unified perspective on resolution and transformation-consistent generative processes.

\section*{Acknowledgement}
This work was supported in part by Institute of Information \& communications Technology Planning \& Evaluation (IITP) grants funded by the Korea government(MSIT) [No.RS-2021-II211343, Artificial Intelligence Graduate School Program (Seoul National University) / No.RS-2025-02314125, Effective Human-Machine Teaming With Multimodal Hazy Oracle Models], the National Research Foundation of Korea(NRF) grant funded by the Korea government(MSIT) (No. RS-2025-02263628), the BK21 FOUR program of the Education and Research Program for Future ICT Pioneers, Seoul National University and Samsung Electronics MX Division.


\appendix
\renewcommand{\thefigure}{S\arabic{figure}}
\renewcommand{\thetable}{S\arabic{table}}
\renewcommand{\theequation}{S\arabic{equation}}
\setcounter{figure}{0}
\setcounter{table}{0}
\setcounter{equation}{0}


\section*{Appendix}

\section{Details on compliance}
We assume compliance of the downsampled trajectory when performing preview generation. A natural question is whether this compliance property genuinely holds. Zhang et al.~\cite{zhang2024flow} provide a proof in the appendix of their work, showing that under an $L$-Rectified Flow, the deviation between the compliant trajectory and the true trajectory is bounded by $O(1/L)$. Their analysis indicates that, although exact compliance is not guaranteed, approximation becomes reasonable once $L$ is sufficiently large. Building on this result, we consider our compliance assumption to be well justified.
\blfootnote{* Authors contributed equally. $\dag$ Corresponding author.}
\section{Detailed Experimental Setup}

\subsection{Experiment compute resources}
All experiments are run on an NVIDIA A100 GPU, which serves as our primary compute resource. Floating operations (FLOPs) were measured using the $\texttt{torch.profiler}$ tool, and the latency results in Tab.~2 are benchmarked directly on the same GPU.

\subsection{Evaluation dataset}
Commonly used evaluation benchmarks such as MS-COCO contain short prompts and limited scene diversity, making them inadequate for representing the aesthetic and complex prompting environments typical of designers or photographers, which are the primary targets of our study. Instead, we adopt the more challenging PixArt-30K evaluation set to better align with real-world usage. From this dataset, we randomly sample 5,000 prompts whose lengths range from a maximum of 1,885 characters to a minimum of 2 characters, with a median of 319 and a mean of 413 characters. This wide spectrum of prompt lengths and expressiveness allows us to capture the dynamic and realistic prompting scenarios encountered in practical workflows.

\subsection{Metrics}
To evaluate the performance of our preview generation, we employ a range of image similarity and image quality metrics. This section provides detailed explanations of each metric, including their computation procedures and specific evaluation criteria.
\paragraph{PIQE~\cite{venkatanath2015blind}.}
PIQE is a no-reference image quality metric that estimates perceptual distortion by analyzing deviations from natural scene statistics at the block level. The image is first transformed into Mean Subtracted Contrast Normalized (MSCN) coefficients, after which each non-overlapping block is classified as either spatially active or uniform based on its MSCN variance. Spatially active blocks are then examined for two types of degradations: noticeable structural distortions, detected via low-variance edge segments, and noise distortions, characterized using a center–surround deviation measure. For each distorted block, PIQE assigns a distortion score derived from the block-level MSCN variance, which is inversely proportional to perceived quality for structural distortions and directly proportional for noise. The final PIQE score aggregates block-level distortions according to
\begin{equation}
    \text{PIQE} = \frac{\sum^{N_\text{SA}}_{k=1}D_{S_k} + C_1}{N_\text{SA} + C_1}
\end{equation}
where $N_\text{SA}$ is the number of spatially active blocks and $D_{s_k}$ is the distortion score of the k-th block. Lower PIQE values indicate higher perceptual quality, making it suitable for evaluating preview fidelity without requiring reference images. We leveraged block size 8, and all other parameters were set to default value.
\paragraph{DreamSim~\cite{fu2023dreamsim}.}
DreamSim is a mid-level perceptual similarity metric learned from the NIGHTS dataset, designed to outperform traditional perceptual metrics such as LPIPS that mainly capture low-level or patch-level distortions. By ensembling CLIP, OpenCLIP, and DINO embeddings and fine-tuning them with LoRA, DreamSim aligns more closely with human visual judgments across variations in pose, layout, shape, and semantic content—factors that LPIPS is not sensitive to. For an image pair $(x, x')$, DreamSim computes similarity using the cosine distance
\begin{equation}
    D(x, x')=1-\cos(f_\theta(x), f_\theta(x'))
\end{equation}
This formulation enables DreamSim to capture perceptual relationships that extend beyond pixel-level fidelity, yielding substantially higher agreement with human preferences compared to LPIPS and other prior perceptual metrics.
\paragraph{DiffSim~\cite{song2025diffsim}.}
DiffSim is a visual similarity metric that leverages the attention features of pretrained diffusion models to overcome the limitations of traditional perceptual similarity measures such as LPIPS, which primarily capture low-level patch statistics and often fail to reflect human judgments of layout, pose, and semantic consistency. DiffSim introduces the Aligned Attention Score (AAS), which aligns the latent representations of two images through the attention mechanism of the Stable Diffusion U-Net and computes a cosine-based similarity on these aligned features. For latent representations $L_A$ and $L_B$ extracted from a chosen attention layer, DiffSim computes
\begin{align}
    \text{AAS}(L_A, L_B)\\
    =\cos(&\text{attn}(Q_A, K_B, V_B),\text{attn}(Q_B, K_A, V_A)), \nonumber
\end{align}
allowing the metric to evaluate both appearance and style similarity while compensating for spatial misalignment. Comprehensive evaluations across human-aligned, instance-level, style, and low-level similarity benchmarks show that DiffSim consistently surpasses LPIPS, CLIP, and DINO, demonstrating substantially higher agreement with human perceptual judgments.
\paragraph{PSNR.}
PSNR is a full-reference image similarity metric that measures the fidelity between a generated image and its reference by comparing pixel-level reconstruction error. It is defined using the mean squared error (MSE) between two images $x$ and $x'$, where a higher PSNR indicates closer pixel-wise correspondence. Formally, for images with maximum possible pixel value $I_{\max}$, PSNR is computed as
\begin{equation}
    \text{PSNR}(x, x') = 10 \log_{10}\!\left(\frac{I_{\max}^2}{\text{MSE}(x, x')}\right).
\end{equation}
Because PSNR is sensitive to exact pixel alignment, it strongly correlates with low-level distortions such as noise, blur, and compression artifacts but is limited in evaluating perceptual or semantic similarity, making it complementary to metrics such as DreamSim, and DiffSim.
\paragraph{FSIM~\cite{zhang2011fsim}.}
FSIM is a full-reference metric designed to capture low-level similarity by focusing on image structures and contrast patterns that are strongly aligned with human visual sensitivity. It relies on two fundamental low-level cues—phase congruency and gradient magnitude—which respectively encode structural information (edges, corners, and local features) and local contrast. Because these cues emphasize fine-grained texture, sharpness, and structural fidelity, FSIM provides reliable measurements of low-level distortions such as blur, noise, compression artifacts, and local geometric degradation. The metric aggregates these cues using a feature-weighted formulation,
\begin{equation}
    \text{FSIM}(x, x') = \frac{\sum_{p} \left( S_{\mathrm{PC}}(p) \cdot S_{\mathrm{GM}}(p) \cdot W(p) \right)}{\sum_{p} W(p)},
\end{equation}
and is widely used when accurate evaluation of pixel-level and structural consistency is required.

\begin{table*}
    \centering
    \caption{The prompts used for the qualitative results in the main paper are ordered from the left column to the right column.}
    \label{tab:prompts_main_quali}
    \begin{tabular}{cc}
    \toprule
    & Prompt used in main qualitative results    \\
    \midrule

    1 & \parbox{0.9\textwidth}{%
        pixelated Quantum entanglement Fentanyl digital banality, dishevelled, avant-garde multicultural diaspora humans contorted jumping twisted broken distorted lighting casts high contrast shadows lighting Hasselblad h6d - 400c, carl zeiss batis super close up 16mm f/ 2.8, ricoh r1
        } \\
    \midrule
    2 & \parbox{0.9\textwidth}{%
        Small Dragon with Finnish elf characters, in a Finnish forest, in the style of korik kokiri, volumetric lighting, lively tableaus, mori kei, dark gold and green, vray, konica big mini. a figure and a mushroom stand among mushrooms in a forest, in the style of rendered in cinema4d, bill gekas, light gold and green, anime-inspired character designs.
        } \\
    \midrule
    3 & \parbox{0.9\textwidth}{
        Chinese man has invented a machine that can travel in time, High and short depth of field, stippling
        } \\
    \midrule
    4 & \parbox{0.9\textwidth}{%
        shadoflectocyberchromos, You got a heart of stone, you can never feel.
        } \\
    \midrule
    5 & \parbox{0.9\textwidth}{%
        Black Hole Sun + ultra high quality + beautiful colours + psychedelic + 3D rendered
        } \\
    \midrule
    6 & \parbox{0.9\textwidth}{%
        A Stone Island X Cottweiler menswear collaboration editorial campaign shot by Johnny Dufort. Pa Salieu is modelling. Metallic fabrics. Mesh inserts. Two-tone materials. Chameleon fabrics. Color-flip materials. Technical treatments and washes to the materials. Night, Raining.
        } \\
    \midrule
    7 & \parbox{0.9\textwidth}{%
        In the image, there are two individuals engrossed in reading a magazine. The person on the left, clad in a black sweater, is holding the magazine with both hands, indicating a deep interest in the content. On the right, another person, wearing a red shirt, is also holding the magazine with both hands, mirroring the actions of the person on the left. They are seated comfortably on a white couch, which stands out against the backdrop of a white wall adorned with a window. The window allows for natural light to filter into the room, illuminating the scene and casting soft shadows. The magazine they are reading is open, suggesting that they are actively engaged in reading it. The pages of the magazine are not visible in the image, but one can infer from their focused expressions that they are reading about something of interest. There is no text visible in the image, and the relative positions of the objects suggest a casual and relaxed atmosphere. The image captures a quiet moment of shared interest and learning between the two individuals.
        } \\
    \bottomrule
    \end{tabular}
\end{table*}

\subsection{Prompts in qualitative results}
Due to space limitations in the main paper, we could not include the prompts used for the qualitative results presented in the main script, and therefore provide them here. They are summarized in \cref{tab:prompts_main_quali}, where the attached indices follow the left-to-right column order. The first four prompts are used for FLUX.1-dev, and the last three prompts are used for the qualitative results of SD 3.5-L. To demonstrate performance under diverse prompting conditions, we deliberately include prompts ranging from short to long.

\section{Additional Experiments}
\subsection{Effect of the number of fixed-point iterations}
In Eq.~(10), the update is defined as a fixed-point iteration over $k$ steps. For efficiency, however, we adopt the setting 
$k=1$ in our main experiments. To assess the effect of using larger values of 
$k$, we additionally evaluate the method with $k>1$.

As shown in Tab.~\ref{tab:supp_k_ablation}, increasing $k$ yields marginal improvement or degradation in performance, but leads to a noticeable increase in runtime. Therefore, using $k=1$ provides a favorable choice for better efficiency.

\begin{table}
    \caption{Quantitative comparison across different values of $k$.}
    \label{tab:supp_k_ablation}
    \centering
    \resizebox{\linewidth}{!}{
    \begin{tabular}{l|cc|ccc}
    \toprule
    $k$ & TFLOPs$\downarrow$ & Speed$\uparrow$ & PIQE$\downarrow$ & DreamSim$\downarrow$ & PSNR (dB)$\uparrow$ \\
    \midrule
    1 & 1178.30 & 1.53$\times$ & 28.55 & 6.83 & 21.182 \\
    2 & 1277.56 & 1.41$\times$ & 28.79 & 6.85 & 21.132\\
    3 & 1376.84 & 1.31$\times$ & 28.88 & 6.76 & 21.132\\
    \bottomrule
    \end{tabular}
    }
\end{table}

\subsection{Preview generation on lower resolution}
In the main paper, we only report preview generation at a resolution of $512\times512$. However, this setting alone is not sufficient to fully assess the effectiveness of the preview generation. Therefore, we additionally conduct experiments at a lower resolution of $256\times256$.

\begin{figure*}[!t]
  \centering
  \includegraphics[width=0.9\linewidth]{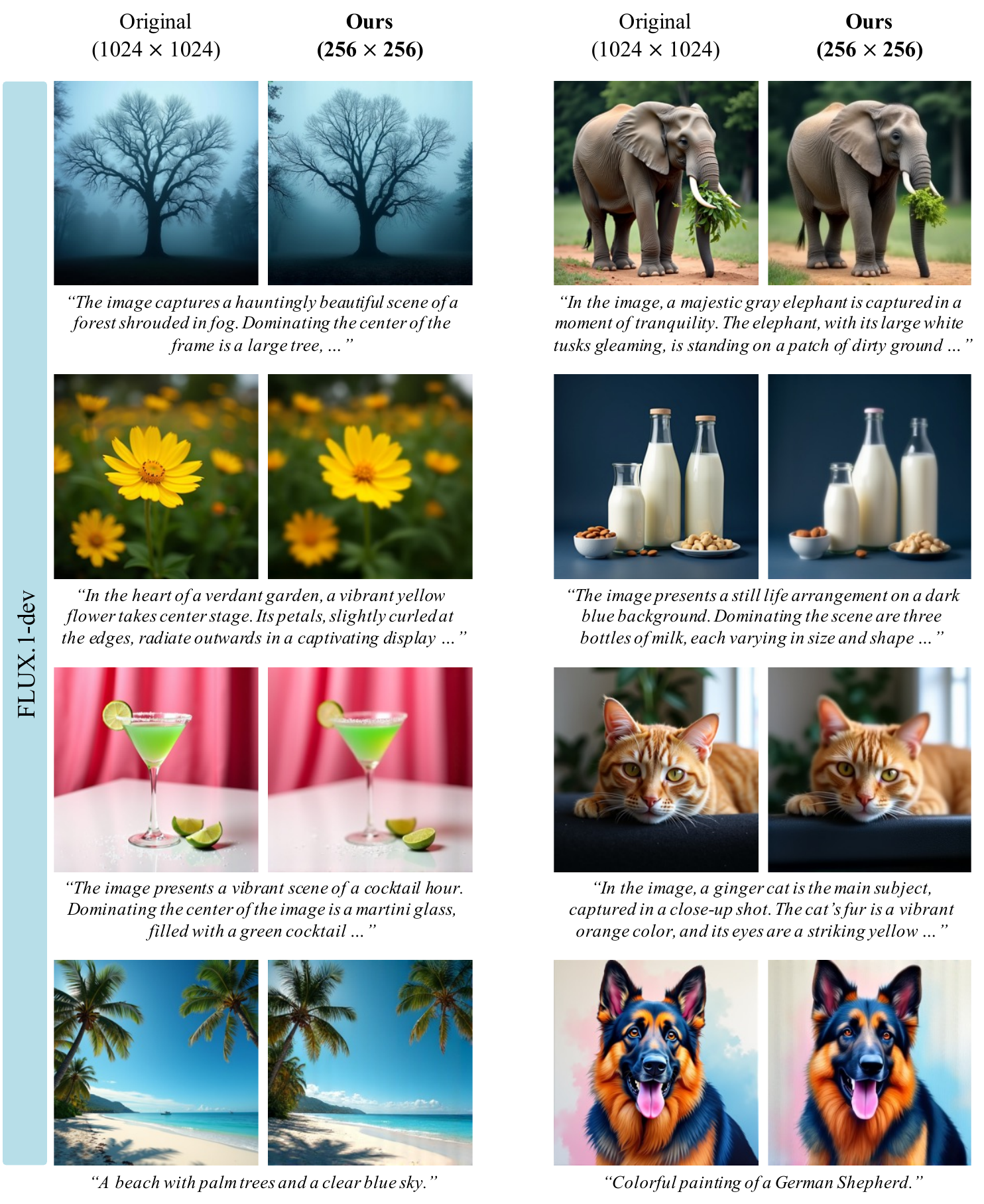}
    \caption{Qualitative results of our proposed method at a lower resolution of $256\times256$.}
    \label{fig:supp_256_quali}
  \hfill
\end{figure*}

Fig.~\ref{fig:supp_256_quali} presents the qualitative results at a resolution of $256\times256$.
While the reduced resolution inevitably introduces slight blurriness, the generated images remain highly perceptually similar to the original images. Although diffusion models exhibit significantly degraded generative capability when directly operating at such a low resolution, the use of commutative-zero guidance effectively mitigates this issue, enabling the production of visually coherent and reliable previews.

\section{Additional Qualitative Results}
This section presents additional qualitative results that could not be included in the main paper due to space constraints.

\subsection{Qualitative Results on FLUX}
\label{sec:supp_quali_flux}
\begin{figure*}[!t]
  \centering
  \includegraphics[width=0.85\linewidth]{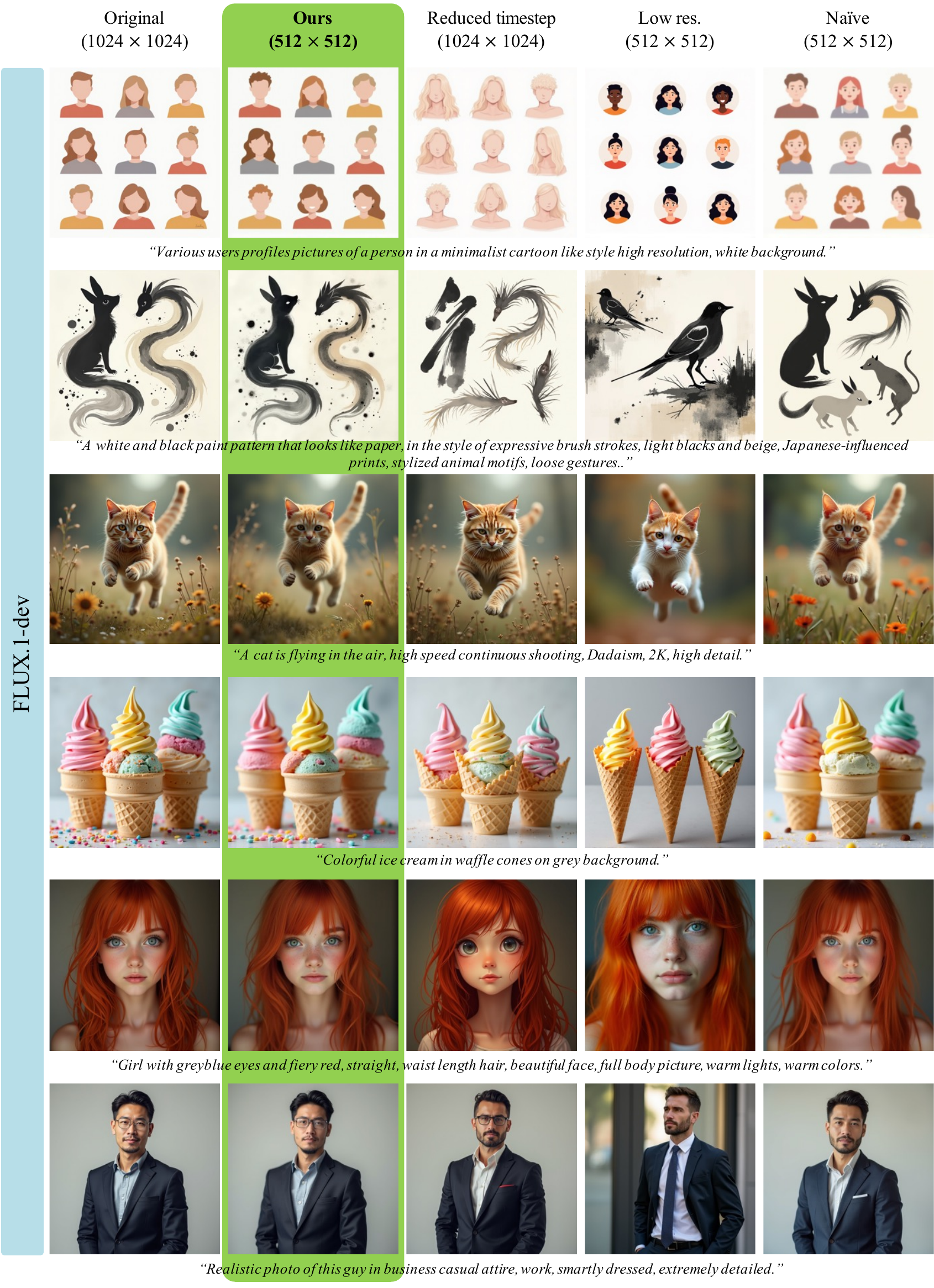}
    \caption{Qualitative comparison of our proposed method on FLUX.}
    \label{fig:supp_flux_quali_1}
  \hfill
\end{figure*}

\begin{figure*}[!t]
  \centering
  \includegraphics[width=0.9\linewidth]{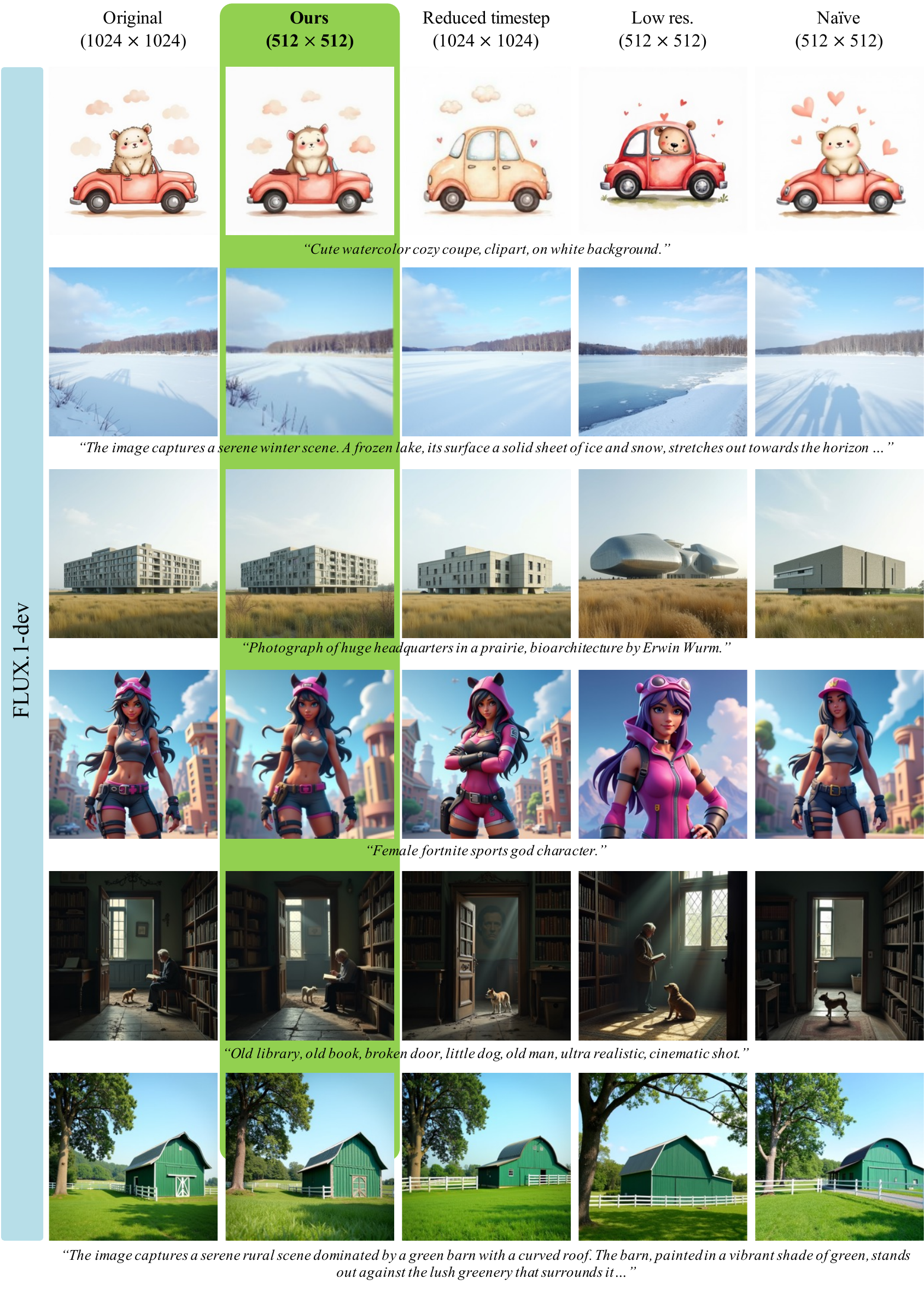}
    \caption{Qualitative comparison of our proposed method on FLUX.}
    \label{fig:supp_flux_quali_2}
  \hfill
\end{figure*}

Fig.~\ref{fig:supp_flux_quali_1},~\ref{fig:supp_flux_quali_2} provide qualitative comparisons demonstrating that our proposed method produces images that are perceptually more similar to the original ones compared with other baselines.

\subsection{Qualitative Results on SD3.5-L}
\begin{figure*}[!t]
  \centering
  \includegraphics[width=0.85\linewidth]{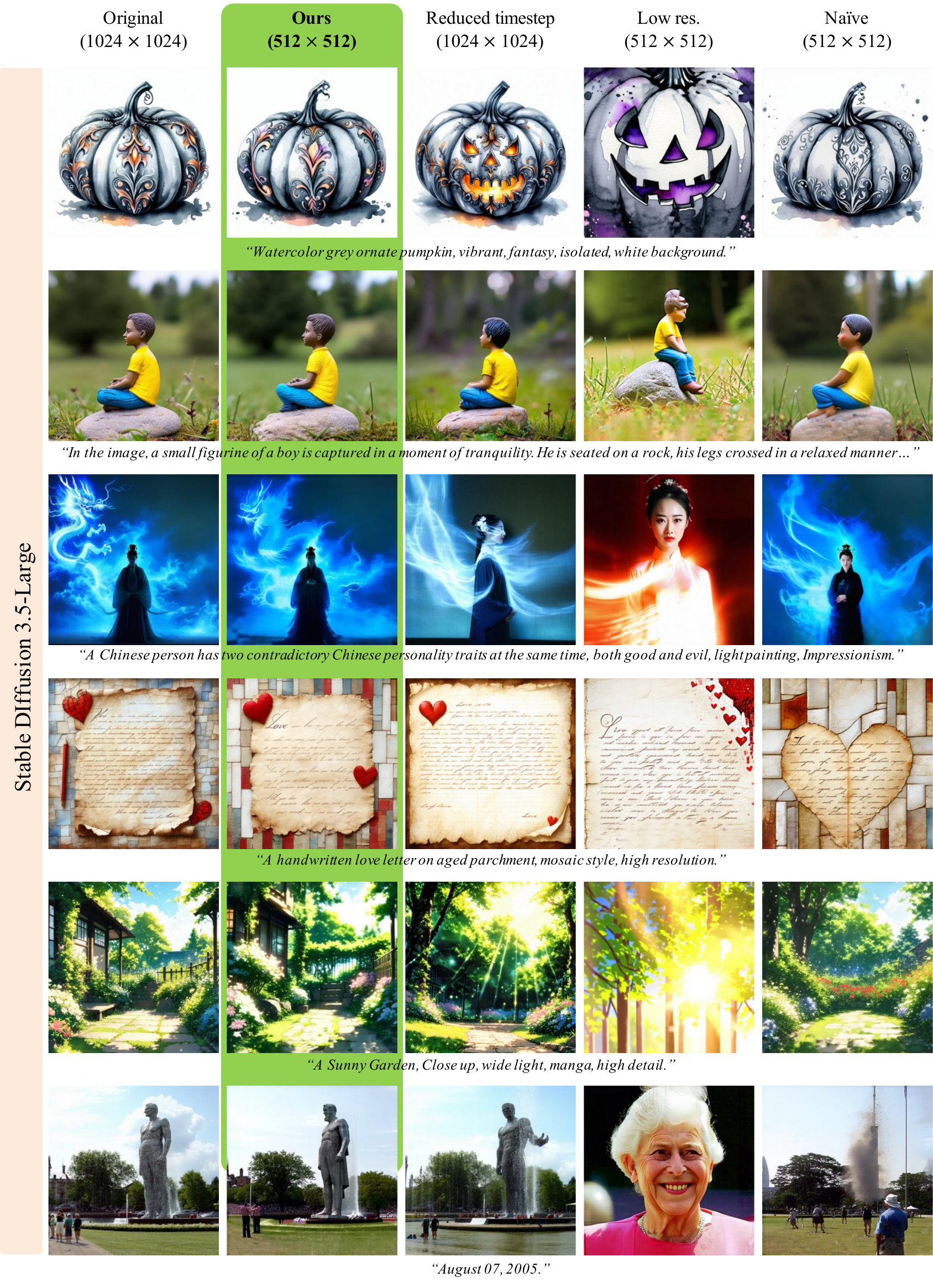}
    \caption{Qualitative comparison of our proposed method on SD3.5-Large.}
    \label{fig:supp_sd3_quali_1}
  \hfill
\end{figure*}

\begin{figure*}[!t]
  \centering
  \includegraphics[width=0.85\linewidth]{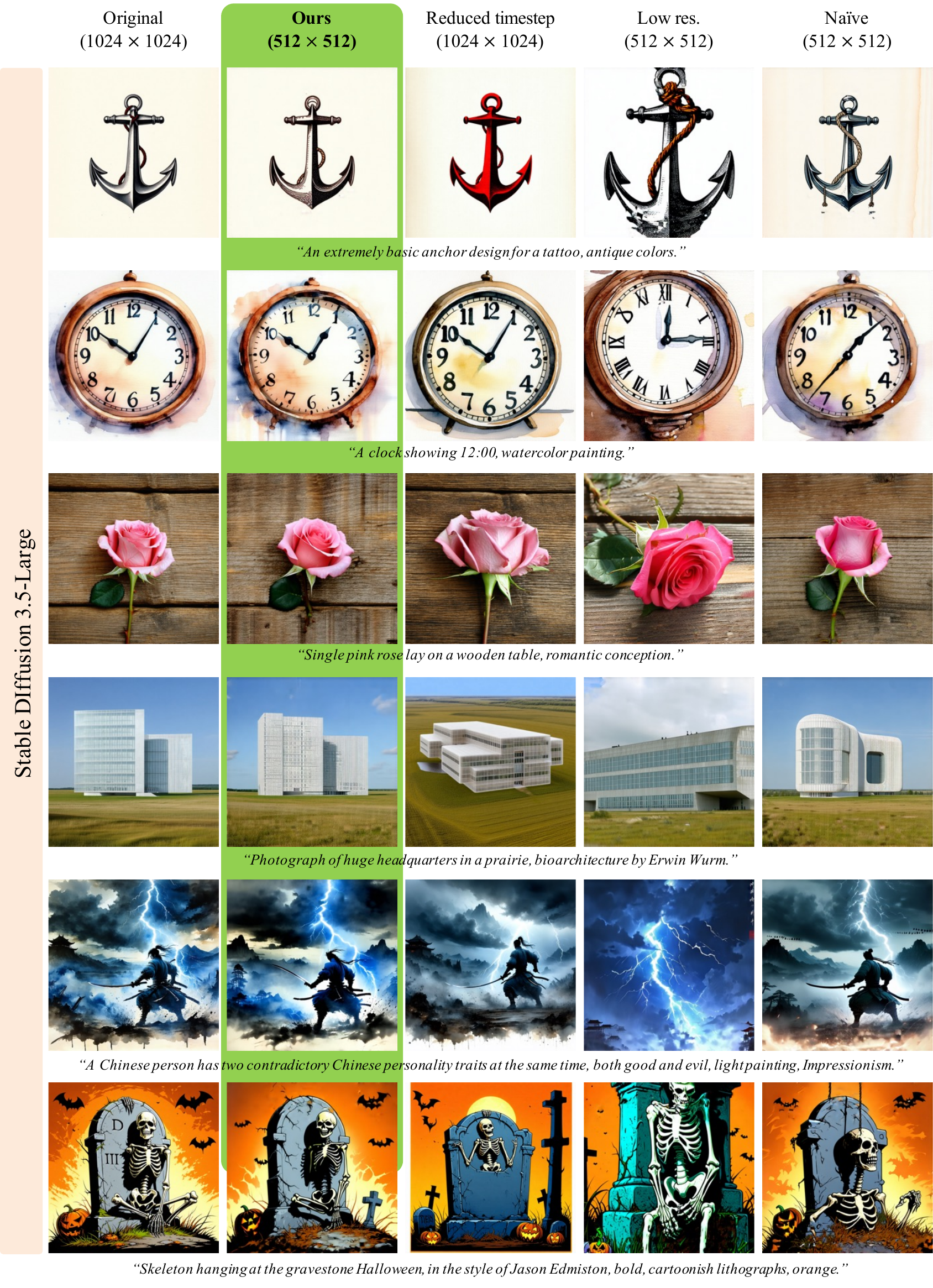}
    \caption{Qualitative comparison of our proposed method on SD3.5-Large.}
    \label{fig:supp_sd3_quali_2}
  \hfill
\end{figure*}

Similarly, Fig.~\ref{fig:supp_sd3_quali_1},~\ref{fig:supp_sd3_quali_2} show that, on Stable Diffusion 3.5 Large (SD3.5-L), our method yields results that better preserve perceptual similarity to the original images than the competing baselines.
\begin{figure*}
    \centering
    \includegraphics[width=\linewidth]{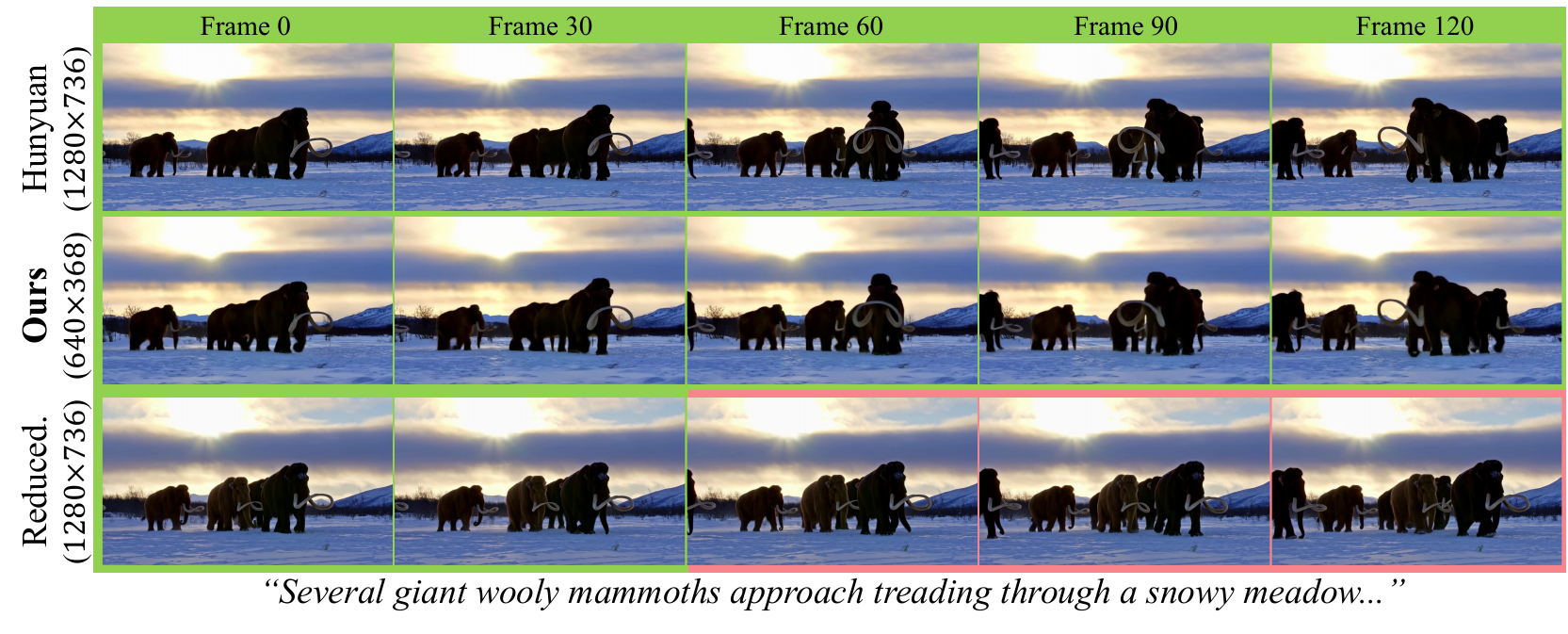}
    \caption{Qualitative results on video diffusion. Using our preview generation framework with HunyuanVideo, we compare against an alternative approach that reduces NFE from 50 to 30. While Mammoth deviates from the full-step composition beyond a certain frame, our preview generation faithfully preserves the composition of the high-resolution content.}
    \label{fig:sup_video}
\end{figure*}
\subsection{Qualitative Results on Video Models}
We conducted video synthesis experiments with HunyuanVideo~\cite{kong2024hunyuanvideo}, as shown in \cref{fig:sup_video}, and observed that the reduced-timestep baseline (NFE 50$\rightarrow$30) yields increasingly inconsistent compositions as the number of frames grows. In contrast, our method generates perceptually consistent LR videos while achieving a $1.75\times$ speedup, with generation times of 1,661 sec (Original), 1,025 sec (Reduced), and 951 sec \textbf{(Ours)} for 120-frame video generation.

\subsection{Integration with Temporal-axis Acceleration}
\begin{figure*}[!t]
  \centering
  \includegraphics[width=0.9\linewidth]{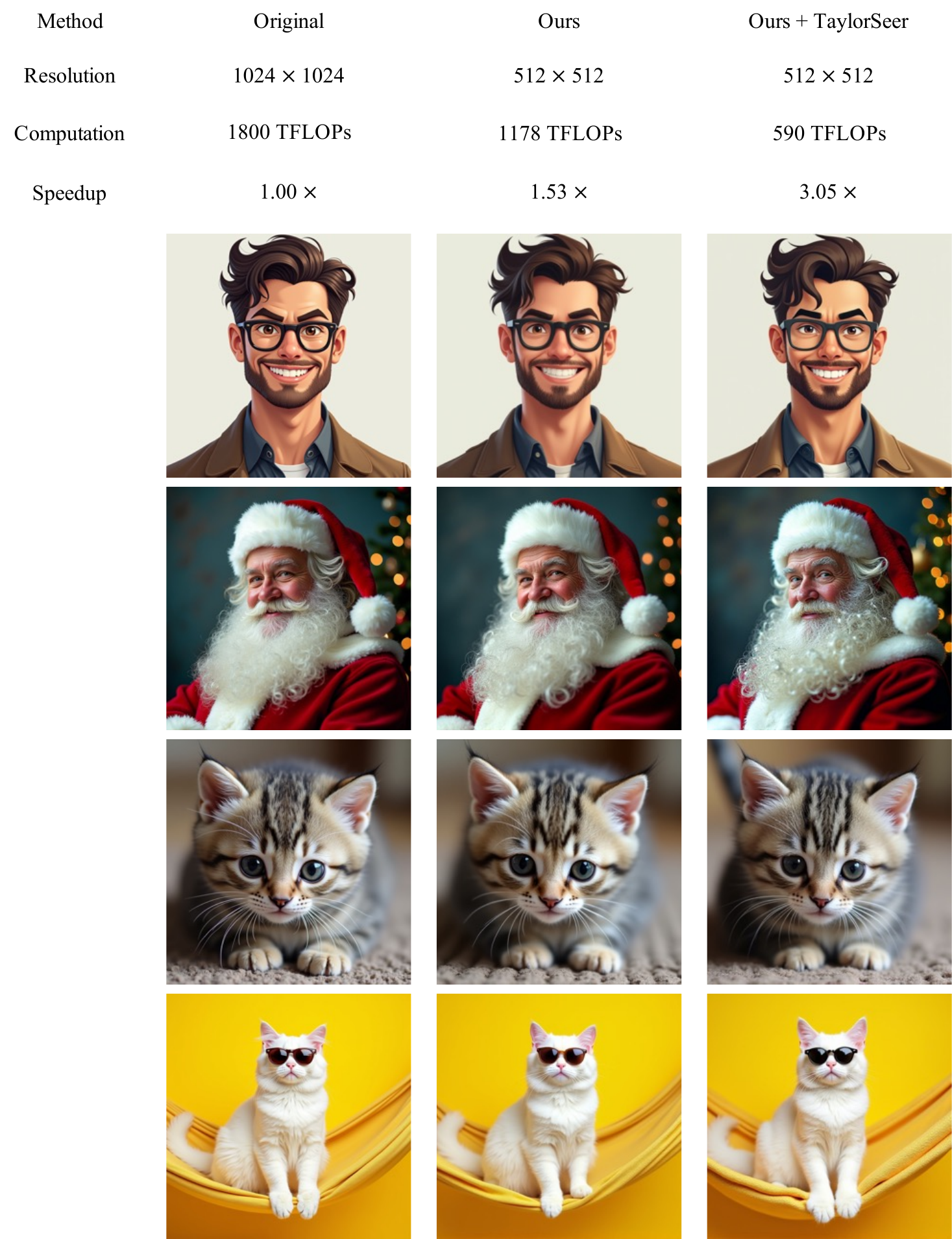}
    \caption{Qualitative comparison after integrating temporal-axis acceleration on FLUX.}
    \label{fig:supp_temporal}
  \hfill
\end{figure*}

Tab.~3 shows that integrating our proposed method with the temporal-axis acceleration technique TaylorSeer~\cite{liu2025reusing} yields additional speedup with almost no performance degradation, improving the acceleration from 1.53$\times$ to 3.05$\times$. Furthermore, Fig.~\ref{fig:supp_temporal} provides a qualitative comparison of the generated images. We observe that the outputs remain nearly indistinguishable both when applying our method alone and when incorporating TaylorSeer, indicating that the original images are preserved with high perceptual fidelity even at the 3.05$\times$ acceleration setting.

\section{Limitation}
The proposed approach relies on the approximate commutator-zero condition and trajectory compliance, both of which hold empirically but are not guaranteed across all architectures or sampling schedules. Since the method operates within a training-free setting, the selected downsampling operators are restricted to mutually exclusive, block-wise matrices, which may limit expressiveness and introduce sensitivity to spatial structure. The reuse of velocity predictions assumes local linearity of rectified flows, which may weaken under models exhibiting strong temporal variation or non-linear behavior, potentially reducing stability for more complex scenes or extreme prompts. Although the method improves LR–HR alignment, the corrections introduce additional computation, presenting an inherent trade-off between accuracy and acceleration. Finally, while the formulation generalizes to certain spatial manipulations, its applicability to broader transformations or non-flow-based generative frameworks remains an open direction for future investigation.

\clearpage
{
    \small
    \bibliographystyle{ieeenat_fullname}
    \bibliography{main}
}

\end{document}